\documentclass[12pt,preprint]{aastex}

\newcommand{\scname}[1]{\textsc{#1}}

\def\ifundefined#1{\expandafter\ifx\csname#1\endcsname\relax}
\ifundefined{ensuremath}\def\ensuremath#1{\relax\ifmmode{#1}}
\else${#1}$\fi\else\relax\fi

\newcommand{\phoenix}{\texttt{PHOENIX}} 
\newcommand{\phx}{\texttt{PHOENIX}}
\newcommand{\synow}{\scname{SYNOW}}

\newcommand{\kms}{\ensuremath{\textrm{km}~\textrm{s}^{-1}}}

\newcommand{\supernovae}{supernov\ae} 
 
\newcommand{\SNIa}{{SN~Ia}} 
\newcommand{\SNeIa}{{SNe~Ia}}

\newcommand{\SiII}{\ion{Si}{2}}

\newcommand{\FeII}{\ion{Fe}{2}}
\newcommand{\FeIII}{\ion{Fe}{3}}

\newcommand{\TiII}{\ion{Ti}{2}}
\newcommand{\SII}{\ion{S}{2}}

\newcommand{\SiIIred}{$\lambda 6355$}
\newcommand{\SiIIblue}{$\lambda 5972$}
\newcommand{\TiIIRSi}{$\lambda 6013$}

\newcommand{\RSi}{$\Re_{Si}$}

\newcommand{\RSiS}{\ensuremath{\Re_{SiS}}}
\newcommand{\RSiSS}{\ensuremath{\Re_{SiSS}}}

\newcommand{\nni}{\nuc{56}\mbox{Ni}}

\newcommand{\nomw}{W7}

\def\ang{\mbox{\AA}}

\ifundefined{nuc}\def\nuc#1#2{\relax\ifmmode{}^{#1}{\protect\textrm{#2}}
\else${}^{#1}$#2\fi}\else\relax\fi

\def\la{\mathrel{\hbox{\rlap{\hbox{\lower4pt\hbox{$\sim$}}}\hbox{$<$}}}}
\def\ga{\mathrel{\hbox{\rlap{\hbox{\lower4pt\hbox{$\sim$}}}\hbox{$>$}}}}
\newcommand{\be}{\begin{eqnarray}}
\newcommand{\ee}{\end{eqnarray}}

\graphicspath{{./}{redone/}}

\shortauthors{Bongard, S. et~al.}
\shorttitle{Spectral Formation in SNe~Ia}

\begin{document}

\title{Multi-layered Spectral Formation in \SNeIa\ Around Maximum Light}

\author{ 
  Sebastien Bongard\altaffilmark{1,2,4},
  E.~Baron\altaffilmark{2,3},
  G.~Smadja\altaffilmark{4},  David
Branch\altaffilmark{2},  and  Peter
H.~Hauschildt\altaffilmark{5}
 }

\altaffiltext{1}{Physics Division, Lawrence Berkeley
  National Laboratory, MS 50A-5014, 1 Cyclotron Rd, Berkeley, CA
  94720-8139 USA; email: {sbongard@lbl.gov}}

\altaffiltext{2}{Homer L.~Dodge Department of Physics and Astronomy, University of
Oklahoma, 440 West Brooks, Rm.~100, Norman, OK 73019, USA; email:
{baron@ou.edu, {branch@nhn.ou.edu}}}

\altaffiltext{3}{Computational Research Division, Lawrence Berkeley
  National Laboratory, MS 50F-1650, 1 Cyclotron Rd, Berkeley, CA
  94720-8139 USA}

\altaffiltext{4}{Institut de Physique Nucl\'eaire Lyon, B\^atiment Paul Dirac
Universit\'e Claude Bernard Lyon-1
Domaine scientifique de la Doua
4, rue Enrico Fermi
69622 Villeurbanne cedex, France; email: {smadja@in2p3.fr}}

\altaffiltext{5}{Hamburger Sternwarte, Gojenbergsweg 112,
21029 Hamburg, Germany; email: {yeti@hs.uni-hamburg.de}}

\begin{abstract}
  We use the radiative transfer code \phx\ to study the line formation
  of the wavelength region 5000--7000~\ang.  This is the region where
  the SNe~Ia defining Si~II feature occurs. This region is important
  since the ratio of the two nearby silicon lines has been shown to
  correlate with the absolute blue magnitude. We use a grid of LTE
  synthetic spectral models to investigate the formation of line
  features in the spectra of SNe~Ia. By isolating the main
  contributors to the spectral formation we show that the ions that
  drive the spectral ratio are \FeIII, \FeII, \SiII\ and \SII. While
  the first two strongly dominate the flux transfer, the latter two
  form in the same physical region inside of the supernova. 
  We also show that the
  na\"ive blackbody that one would derive from a fit to the observed
  spectrum is far different than the true underlying continuum.
\end{abstract}

\keywords{cosmology --- stars: atmospheres --- supernovae} 


\section{Introduction}

Type Ia \supernovae\ have been used as ``standardizable candles'' for
more than 10 years, thanks to the correlation between their light
curve shape and their absolute blue magnitude at maximum light. This
correlation is well matched by a spectroscopic sequence
\citep{nugseq95} defined by the ratio, \RSi, of the depth
of two absorption features usually identified as \SiII\  \SiIIblue\ and
\SiIIred\ lines. Using the radiative transfer code \phoenix\
\citep[][and references therein]{hbjcam99,bhpar298,hbapara97,phhnovetal97,phhnovfe296},
this sequence has been matched 
to a temperature sequence that could physically be related to the
amount of \nni\ produced in the context of Chandrasekhar-mass white
dwarf explosions.   

\citet{hathighv00} showed that \RSi\ and the velocity derived from the
minimum of the $\approx 6100$~\AA\ feature correlated poorly,
suggesting that a one-parameter description of \SNeIa\ was
insufficient.  This velocity derived from the P-Cygni feature
associated with the \SiII\ \SiIIred\ line, and its time evolution
break \SNeIa\ into subclasses depending upon more than one parameter
\citep{benetti05}.  \citet{branchcomp206} and \citet{hach06} also
showed that different spectroscopic indicators exist which can be used
to distinguish variations even within the ``normal'' \SNeIa.  In the
context of Chandrasekhar-mass explosions, these variations cannot be
interpreted only in terms of differences in the amount of \nni\
produced.  The velocity scatter could be due to properties of the
progenitor system that impact on the kinetic energy released during
the explosion. For example, the ratio ($^{54}$Fe+$^{58}$Ni)/$^{56}$Ni
has been proposed as a second physical parameter which could account
for this dispersion \citep{mazzpod06}.

This paper is an attempt to unravel the spectral formation in the
 $5000-7000$~\AA region (henceforth the \RSi\ wavelength region),
 as a way to address the  
physical reasons for their correlation with luminosity, and especially
the link between this correlation and the temperature sequence of
\SNeIa.
A previous attempt using  the radiative transfer code \synow\
\citep{fisher00}, was performed by \citet{garn99by04} who explained  
the counter-intuitive correlation of \RSi\ with luminosity by the
temperature sensitivity of a \TiII\ line. 
On the other hand
synthetic spectral investigations  \citep{stehle02bo05,branchcomp206}
failed to find any important \TiII\ line contributions in this part of
the spectrum.

We use the radiative transfer code \phx\ differently than it has
usually been used previously. Instead of fitting a supernova by varying the time
and bolometric luminosity, we use a grid of simulated models to probe
the line formation process. Our input model is the parametrized
deflagration model W7 \citep{nomw7}, homologously expanded to
calculate the abundances at any given time.  The knowledge of the
abundance structure of the explosion model, and the self-consistent
physical structure provided by \phoenix\ are used to describe more
precisely how the spectrum is formed in \SNeIa.  \citet{bongard06a}
introduced the line ratio \RSiS, which is defined as the ratio of two
integrals of the flux over a chosen wavelength range, where each integral
is centered around the two lines used in the original definition of
\RSi\ (see \S\ref{rsisdef}). It is designed to be useful for spectra
which have lower signal 
to noise than well-observed nearby SNe~Ia.

This understanding of the formation of the spectrum allows us to
explain why the line ratio \RSiS\ is a good probe of the temperature
sequence, and reinforces the link between this sequence and the
absolute blue magnitude of \SNeIa.  These results also allow us to
rule out a \TiII\ contribution to the correlation of \RSi\ with
luminosity.  We emphasize that even though the synthetic spectra
presented here \emph{do not} do a good job of reproducing observed
SNe~Ia spectra, we consider W7 to be the fiducial SNe~Ia model.
\citet{bbbh06} showed that reasonably good synthetic spectra can be
produced with W7 near maximum light. We make differential comparisons
based on that model, keeping in mind that \citet{bbbh06} emphasized
that W7 did not reproduce the wavelength region that we examine in
detail here.


We will address time
evolution only qualitatively. 
The goal of this paper is to understand the temperature sequence of
SNe~Ia spectra around maximum light, thus we concentrated on day 20 
after explosion, since maximum light in $B$ in SNe~Ia has been shown  
to occur 18--25 days after explosion.

\section{Analysis tool: \phx\ simulations}
\subsection{The LTE grid}
 
Using the multi-purpose radiative transfer code \phx\ we converged LTE
calculations for the \nomw\ explosion model at 10, 15, 20, and 25 days
after explosion, for a range of bolometric luminosities spanning the
``normal'' \supernovae\ blue magnitudes.

\subsection{Single Ion spectra\label{subsec:sing_ion}}

Each model converged with \phx\ not only provides a spectrum, but also
the knowledge of the physical structure and line optical depth in the
supernova. This gives us the ability to compute ``single ion 
spectra'' as follows: Using the converged output of the \phx\
simulation, we artificially  turn off all but the
continuum and one single ion line opacities. The
``continuum opacities'' denote all the bound-free and free-free
opacities as well as electron and Rayleigh
scattering.  We then recalculate the solution of
the scattering problem in order to get the ``single ion
spectrum'', with the level populations and free electron number kept
fixed.

The relative strength of the absorption and emission features in this
spectrum give an indication of the contribution of each line to the
complete spectrum. Since the solution of the scattering
problem is recalculated, the features are \emph{not} expected to look
exactly the same as in the complete spectrum.  In any case, due to the
global nature of the transfer equation, the
full spectrum in \emph{not} the linear combination of all the single
ion spectra.

In these single ion spectra, the flux transfer from the blue
to the red due to scattering will be smaller, due to  the opacity
decrease caused by the suppression of most of the ions. They will
thus appear bluer than the complete spectrum. The back scattering will
also be decreased for the same reason, resulting in ``single ion
spectra'' brighter than the full synthetic spectrum. 

The ratio of scattering to absorption is larger in the single
ion spectra compared to the full synthetic spectrum because the
full scattering problem is solved with fixed level populations and
fewer atomic lines. The same effect enhances stimulated emission in the
single ion spectra, resulting in strong net emission in the
P-Cygni profiles.

Since the single ion spectra are always brighter than the
full synthetic spectrum we must normalize the luminosity in order to
compare the two spectra. We do this by setting the single ion
bolometric luminosity to that of the full synthetic spectrum. This
procedure is
trivial since the synthetic spectra include fluxes at all wavelengths.

The special case we name ``continuum-only spectrum'' is that where all
line opacities are set to zero, and therefore the spectrum is due to
pure continuum processes.  In a moving atmosphere a
photoionization edge can produce a feature that looks like a P-Cygni
line \citep{b94i2}. We can turn on and off different
species individually, thus we can, for example, have a ``two-ion''
spectrum that consists of, for example, only Si~II and S~II.

\section{Using single ion spectra: ruling out a \TiII\
  contribution to \RSi}

\subsection{The \RSi\ puzzle }

\begin{figure}[ht]
\centering
\includegraphics[width=0.70\textwidth, clip]{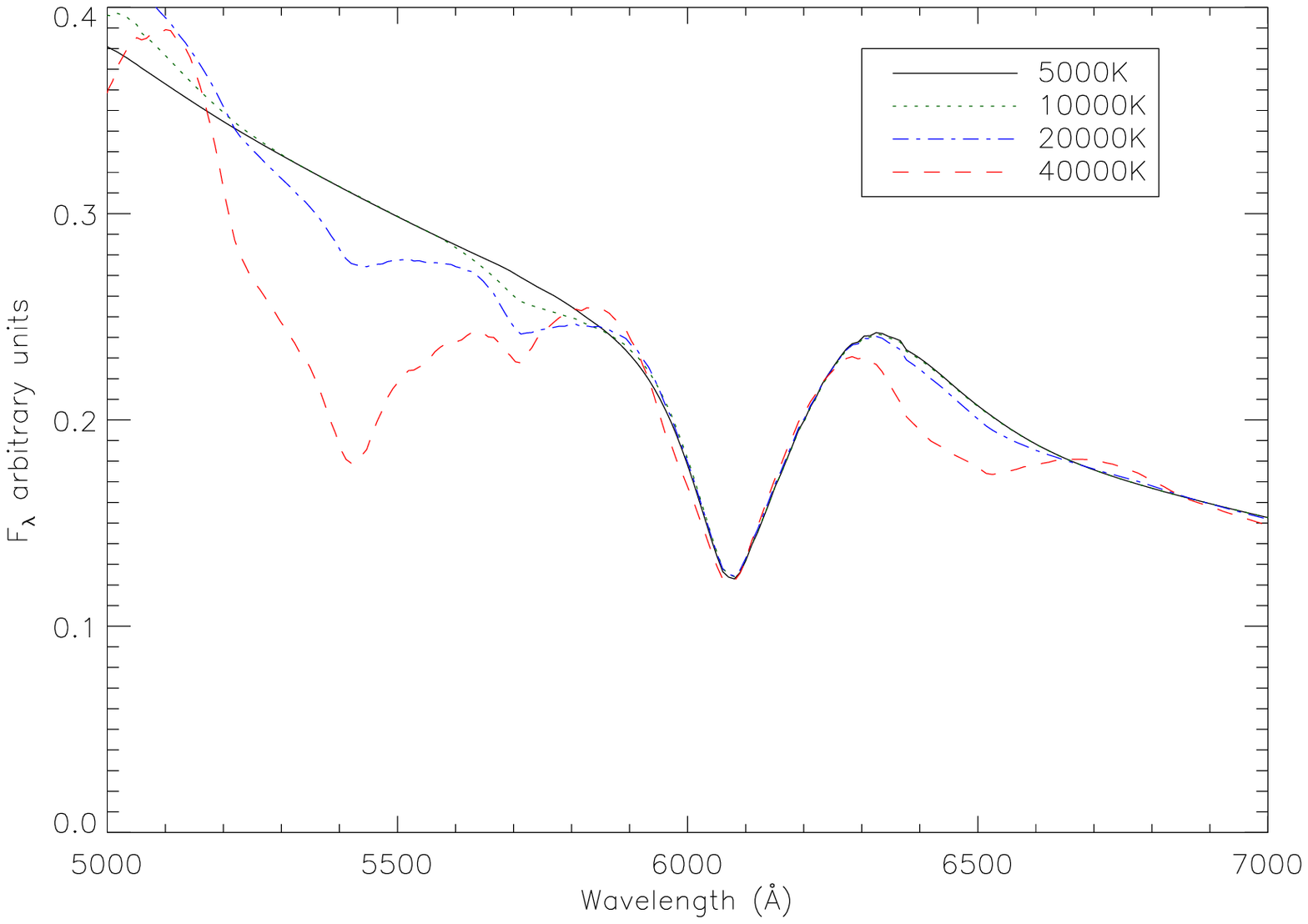}
\caption{\SiII\ in \synow\ for increasing excitation 
  temperatures.
  The trough of the blue feature due to \SiII\
  \SiIIblue\ increases relative to the trough of the
  \SiII\ \SiIIred\ feature 
  as temperature increases, as predicted by
  Eq.~\ref{eq:SiTauRedBlue}. The \SiII\ reference line is \SiII\
  \SiIIred\, and only the excitation temperature is varied, which
  explains the $\sim 6100$~\AA\ trough stability. (Note that the
  unrealistic excitation temperature of 40000K leads to a very strong
  feature around 5400~\AA\ that is not observed in SNe~Ia spectra.)
} 
\label{fig:RSiseqTex}
\end{figure}

Because of their similar redshift with respect to \SiII\ \SiIIred\ and
\SiII\ \SiIIblue\ lines it is tempting to consider the $6100$~\AA\
and  $5800$~\AA\ troughs to be part of their P-Cygni  profiles. These two
lines share the 4P level, the \SiIIred\ line is the 
4S-4P transition and the \SiIIblue\ is the 4P-5S transition. Their optical
depths can thus be estimated in the Sobolev approximation using \citep{atlas99}:
\begin{equation}
  \label{eq:tauSi}
  \tau=(\frac{\pi e^{2}}{m c})f\, \lambda \, t \, n_{l}\, [1-\frac{g_{l}
      n_{u}}{g_{u} n_{l}}],
\end{equation}
where $n_{l}$ and $n_{u}$ are the number densities
of the lower and upper level associated with the transition, $f$ is its
oscillator strength, $t$ is the time since explosion, $g_{l}$ and
$g_{u}$ are the statistical weights of the levels,  and $\lambda$ is the
wavelength of the transition. Assuming thermal equilibrium, the 
ratio of the \SiII\ \SiIIred\ line optical depth ($\tau_{red}$) over
the \SiII\ \SiIIblue\ line optical depth ($\tau_{blue}$)  becomes: 
\begin{equation}
  \label{eq:SiTauRedBlue}
  \frac{\tau_{red}}{\tau_{blue}} \propto
  \frac{\frac{1}{3}~e^{\Delta_{E_{red}}/(k_{B}T)}-1}{1-3~e^{-\Delta_{E_{blue}}/(k_B
      T)}}
\end{equation}
where $\Delta E$ is the energy difference between the upper and
lower levels of the transition.
This ratio decreases monotonically with respect to a
temperature increase, as also shown in Fig.~\ref{fig:RSiseqTex},
where \synow\  was used to compute \SiII\ spectra for excitation
temperatures ranging from 5000~K to 40000~K.

Since P-Cygni troughs become deeper with increasing optical depth, a  
higher \SiII\ excitation temperature should increase \RSi. Under the
reasonable assumption that  higher luminosity \SNeIa\ have
higher \SiII\ excitation temperatures, \RSi\ should increase and    
not decrease with luminosity. Observations show the opposite
behavior. 

\clearpage
\clearpage

\begin{figure}[ht]
\begin{center}
\includegraphics[width=0.70\textwidth, clip]{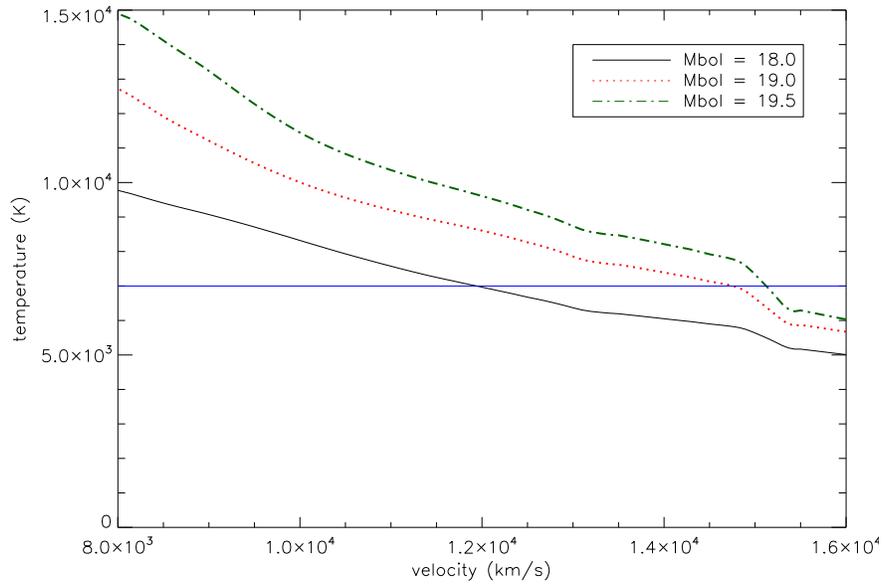}
\caption{Temperature 
for day 20 models in the $8000$-$16000$~\kms\ region
where \SiII\ ions are found, as a function of the model parameter
$M_{bol}$. The horizontal blue line marks the 7000 K limit at which
\TiII\ lines have been suggested to dominate over \SiII\ \SiIIblue\ 
\citep{garn99by04}.}
\label{fig:RSiTempComp}
\end{center} 
\end{figure}

Fig.~\ref{fig:RSiTempComp} displays the temperature structure from
our grid of calculations 20 days after explosion with increasing
bolometric luminosities. We restricted the plot to the
$9000-16000$~\kms\ region, where silicon is found in W7. This shows the
physical temperature increases with bolometric luminosity as one
would expect. In LTE, a temperature increase results in a \SiII\
excitation temperature 
increase with luminosity which would lead to an \RSi\ increase with
luminosity. Therefore, unless one assumes a temperature 
structure inversion in the $9000-16000$~\kms\ region due to NLTE
effects, \SiII\ lines alone are unable to explain the trend of the
\RSi\ correlation with \SNeIa\ blue magnitude. 

\clearpage
\clearpage

\subsection{\TiII\ impact on the ``\RSi\ wavelength region''}

\citet{garn99by04} proposed that \RSi\ variation with luminosity could be
accounted for by a blend of  \TiII\ lines. We focus on the
behavior of the \TiIIRSi\ line. \citet{garn99by04} suggested that the
the \RSi puzzle, described above could be solved by the
temperature 
evolution of this line, which increases much faster than the \SiII\
\SiIIblue\ line decreases below 7000~K. 

\begin{figure}[ht]
  \centering
  \includegraphics[width = 0.7\textwidth, clip]{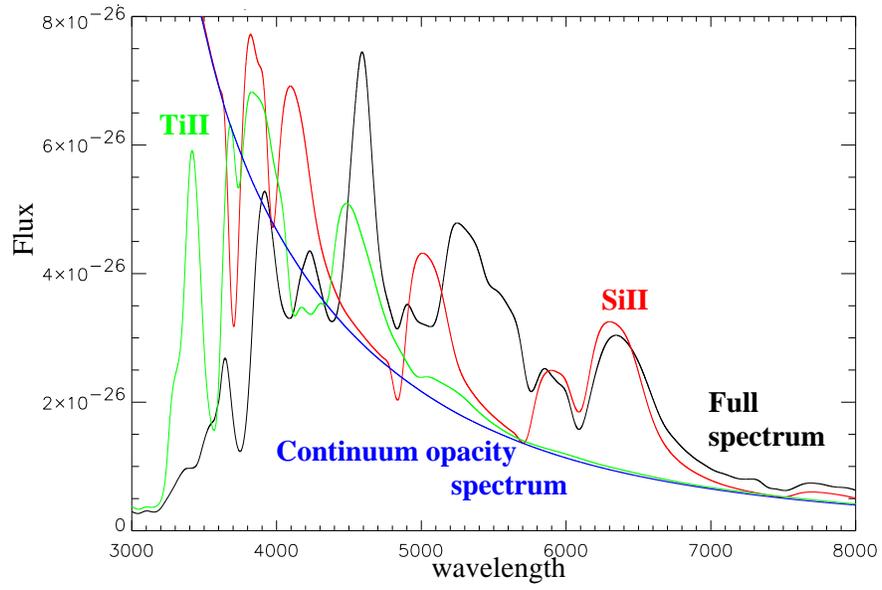}
  \caption{The single ion day 20 spectrum of \SiII\ and \TiII\ alone as
    well as the full spectrum for our lower luminosity model. We also
    display the continuum-only spectrum as a
    reference.}
  \label{fig:phxTiIISiII}
\end{figure}

In order to probe this assumption we display in
Fig.~\ref{fig:phxTiIISiII} the day 20 two-ion spectra of \SiII\ and
\TiII\ alone as well as the full spectrum for our lower 
luminosity model. 
We also display the continuum-only spectrum as a reference. 
For \SiII, the 6100~\AA\ trough and the concomitant emission peak
are, as expected, dominated by \SiII\ lines, but the  \SiII\ \SiIIblue\
P-Cygni profile lacks the full spectrum blue edge, hinting at a
missing contribution to the full spectrum line profile.  
This other contribution can not be \TiII\ as no \TiII\ lines appear in
the \RSi\ wavelength region, even though the \TiII\ \TiIIRSi\ line has
been checked to form between 12000 and 16000~\kms, where the
temperature is below 7000~K as indicated by the black line in
Fig.~\ref{fig:RSiTempComp}.   

One could argue that this line does not appear in our two-ion
spectrum because of a too low abundance of titanium in \nomw. We rule
out the low abundance in W7 as an explanation for this effect since
the \TiII\ single ion spectrum shows a strong feature to the blue of
5000~\AA, which does a reasonably good job of reproducing the observed trough
seen in fast decliner SNe~Ia such as SN~1991bg.

\clearpage
\clearpage

Fig.~\ref{fig:RSiTempComp} shows the \phoenix\ temperature structure
between 10000 and 16000~\kms, where the \SiII\ and \TiII\ lines of the
\RSi\ wavelength region form in all our synthetic spectra models.
Using \synow\ we were able to fit the \RSi\ wavelength region with
\SiII\ and \TiII\ lines alone, but required a \TiII\ excitation temperature
of 40000~K. This temperature is way too high compared to the  detailed
\phx\ results, but note that the temperature in \synow\ is dependent
on the chosen reference line. However, even with \synow, one
expects that \TiII\ line strength increases with \emph{decreasing}
temperature. What is happening in this case, is that in order to
obtain a noticeable contribution of \TiII\ (with respect to \SiII), we
must go to very high excitation temperatures combined with extremely
large optical depths in \TiII: the 
temperature dependence is not independent of the \TiII\ optical depth,
which in \synow\ is a free parameter.

\begin{figure}[ht]
  \includegraphics[width = 0.47\textwidth, clip]{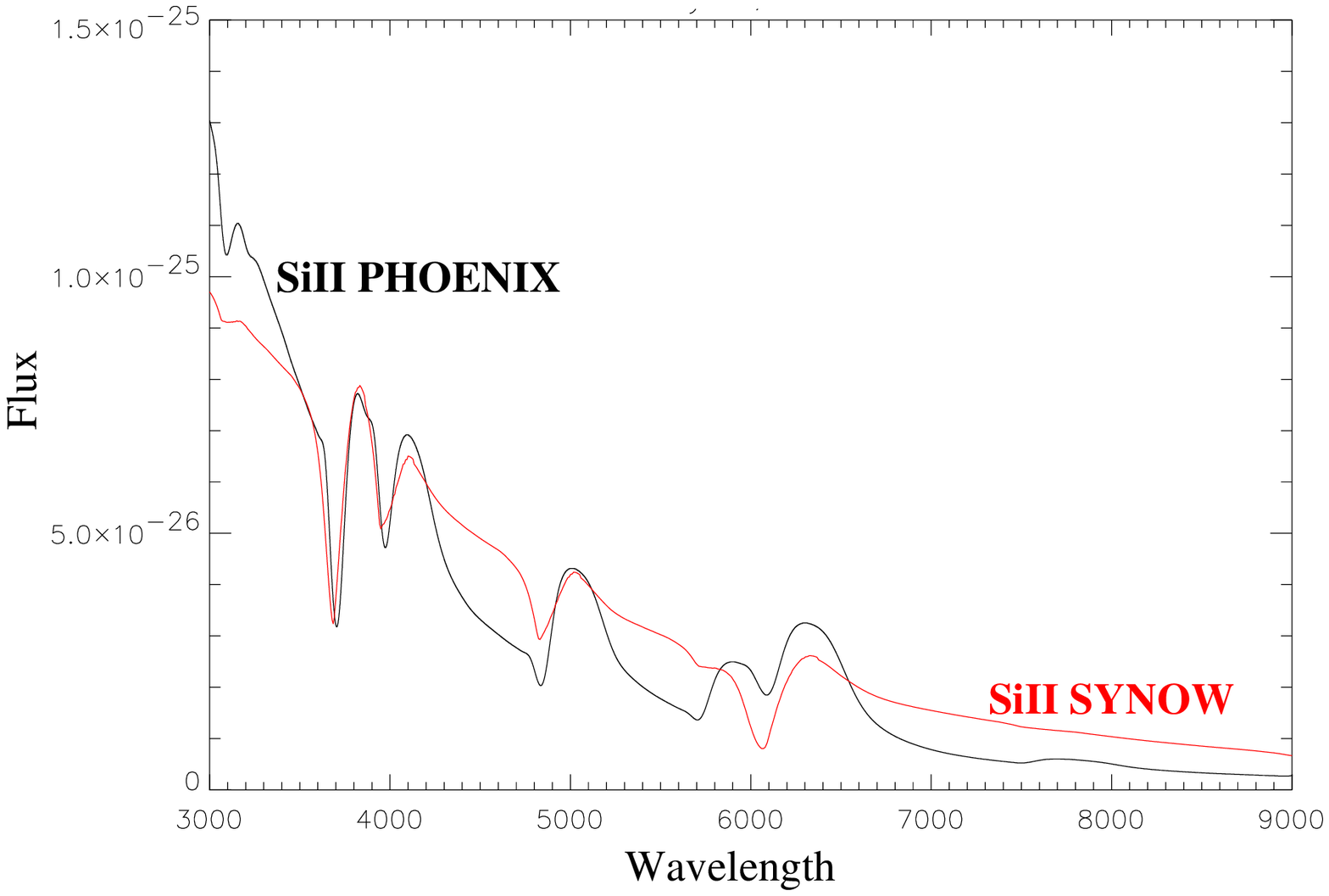}
  \includegraphics[width=0.47\textwidth]{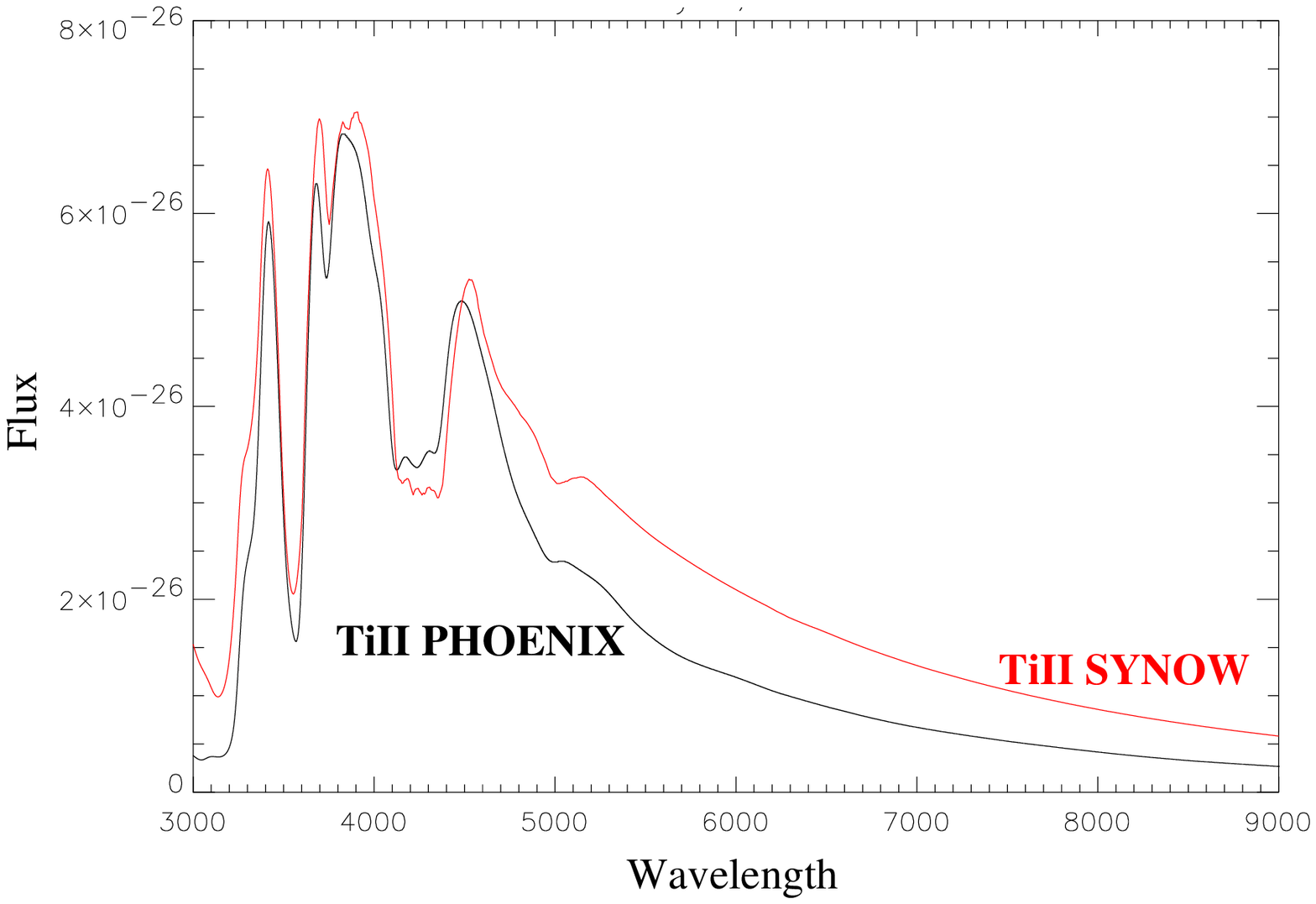}
  \caption{\synow\ and \phoenix\ single ion spectra 
    with \synow\ excitation temperature
    matching \phoenix\ temperature structure. Left panel: \SiII, right
  panel \TiII.}
  \label{fig:SiTiSynowPhx}
\end{figure}

This very high excitation temperature does not necessarily indicate that the
assumptions of \synow\ have broken down, since the \phoenix\ and
\synow\ \TiII\ spectra agree quite well for similar temperatures. 
In Fig.~\ref{fig:SiTiSynowPhx} we display the comparison between the
\synow\ results and the \phx\ results in the temperature range
appropriate to the physical results from \phx.  The 
\SiII\ and the \TiII\ line strengths agree.  We also agree with the point of
\citet{garn99by04} that the temperature dependence of the 
strength of \TiII\ is such that the line-strength increases
dramatically below 7000~K; however, the  relative strength of the line
with respect to \SiII\ does not increase fast enough for the
\TiII\ line to become important at reasonable physical
temperatures. This is due to the fact that the initial line
strength of \TiII\ is so small.
\clearpage
\clearpage

\section{Multi-layered spectrum formation}

\subsection{The deepest layer}

The multi-layer spectrum formation picture alters the way that one
thinks about how spectrum formation occurs. In the photospheric
picture, one thinks about features forming in a reversing layer, above
a true continuum. In the multi-layer picture, features can form
throughout the supernova atmosphere and can in principle imprint a
shape into the ``continuum'' that is altered higher up in the
atmosphere. The most important difference between the multi-layer
picture and the photospheric picture is that features from multiple
ionization stages can strongly affect the overall spectrum.

\clearpage
\begin{figure}[ht]
\centering
  \includegraphics[width = 0.45\textwidth,clip]{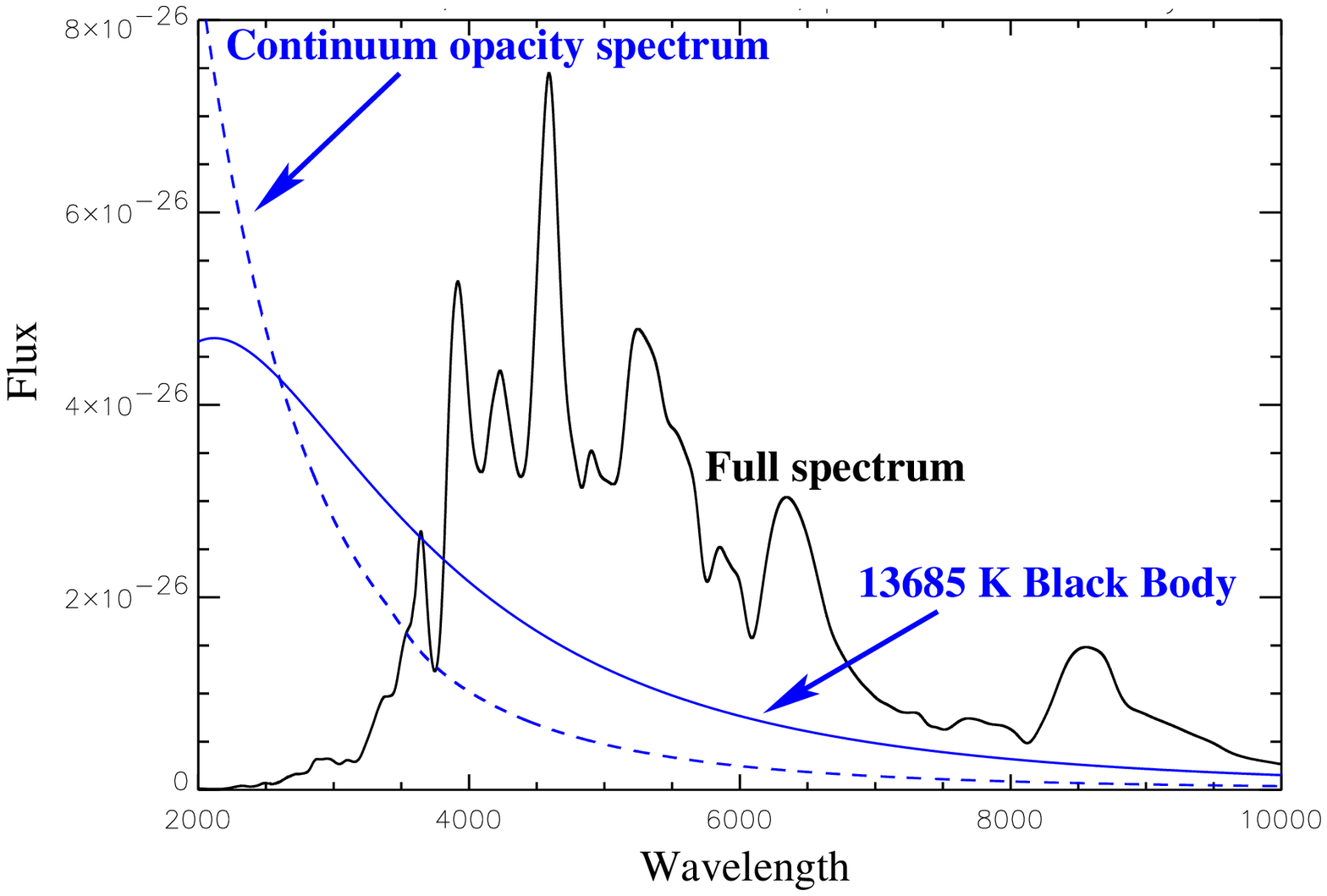}\\
  \includegraphics[width = 0.45\textwidth,clip]{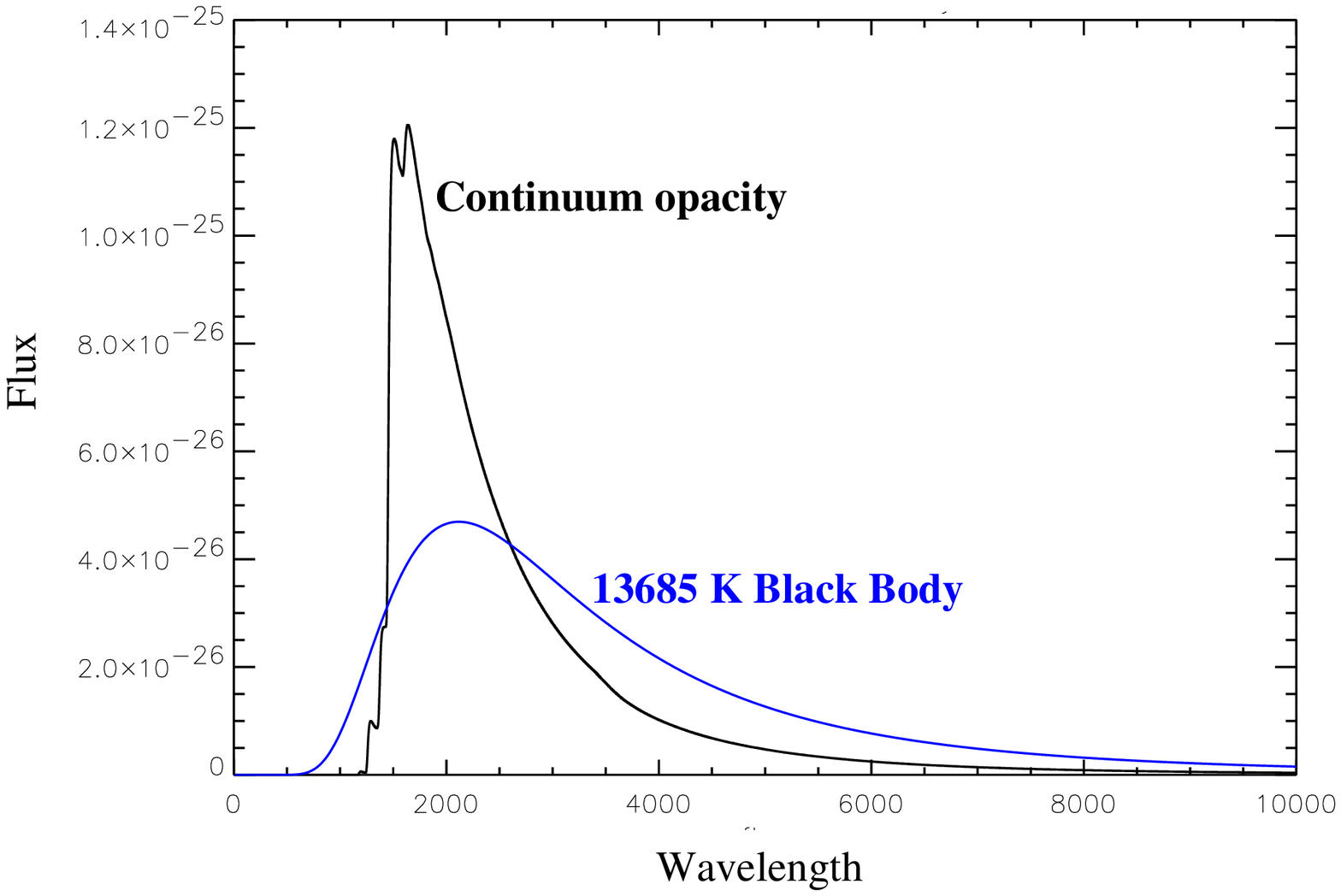}
  \caption{The na\"ive blackbody fit would peak where the full
    synthetic spectrum does, that is around 5000~\ang, far to the red
    of the physical blackbody. 
    Top Panel: Full \phx\ day 20 spectrum in black,
    continuum-only spectrum forming below 2000~\kms\ in dotted blue,
    $13685K$ blackbody in plain blue. Bottom Panel: Continuum-only day 20
    spectrum in dotted blue, and $13685K$ blackbody in plain
    blue.}
  \label{fig:phxOnlyNoone}
\end{figure}
\clearpage

In Fig.~\ref{fig:phxOnlyNoone} (top panel) we display the deepest layer of the 
spectral formation of our lower bolometric luminosity model: the
continuum-only spectrum.
It forms below 3000~\kms\ for all our range
of bolometric luminosities, where the continuum optical depth becomes
greater than one,  which is at a much
lower velocity than where the photosphere is usually considered to be
in the  photospheric model.

Figure~\ref{fig:phxOnlyNoone} (bottom panel) clearly shows that the na\"ive
blackbody fit (that is the blackbody which peaks at the same
wavelength as the smoothed full synthetic spectrum) to the full
synthetic spectrum would give a blackbody temperature that is much
cooler than the true underlying continuum. While the  
continuum-only spectrum is brighter and bluer (as described in
\S\ref{subsec:sing_ion}) Fig.~\ref{fig:phxOnlyNoone} (bottom
panel) shows that 
the blackbody obtained from the physical temperature 13685~K is also
significantly bluer than the na\"ive blackbody. This is a very
important point since it shows that the na\"ive blackbody has nothing
to do with the real physical conditions.

It is crucial to keep this point in mind when analyzing SNe~Ia
features, especially when using techniques like principle component
analysis \citep[PCA,][]{james_pca06}.
It is tempting to subtract a ``continuum'' case by
case in order to maximize the contribution of the lines. Since    
the ``pseudo-continuum'' is responsible for a large part of the 
spectral structure, this procedure yields the opposite result: the
subtracted  ``continuum'' adds a fake relationship between features. 
The PCA eigenvectors will thus no longer account only for real SNe~Ia
diversity, but also for the arbitrary continuum.

The blackbody of Fig.~\ref{fig:phxOnlyNoone} (bottom panel)
corresponds to the temperature where $\tau_{e} \sim 1$. Even at depth
of 2000~\kms\ ($\tau \sim 3 $ at 2000~\kms), the supernova
spectrum is \emph{not} a blackbody spectrum. The spectrum of the full
\nomw\ model is obtained from the steep narrow blue continuum-only spectrum
by line interactions that transfer flux toward the red by Doppler
shift or fluorescence.

\subsection{Iron lines}

\begin{figure}[ht]
  \centering
  \includegraphics[width=0.48\textwidth]{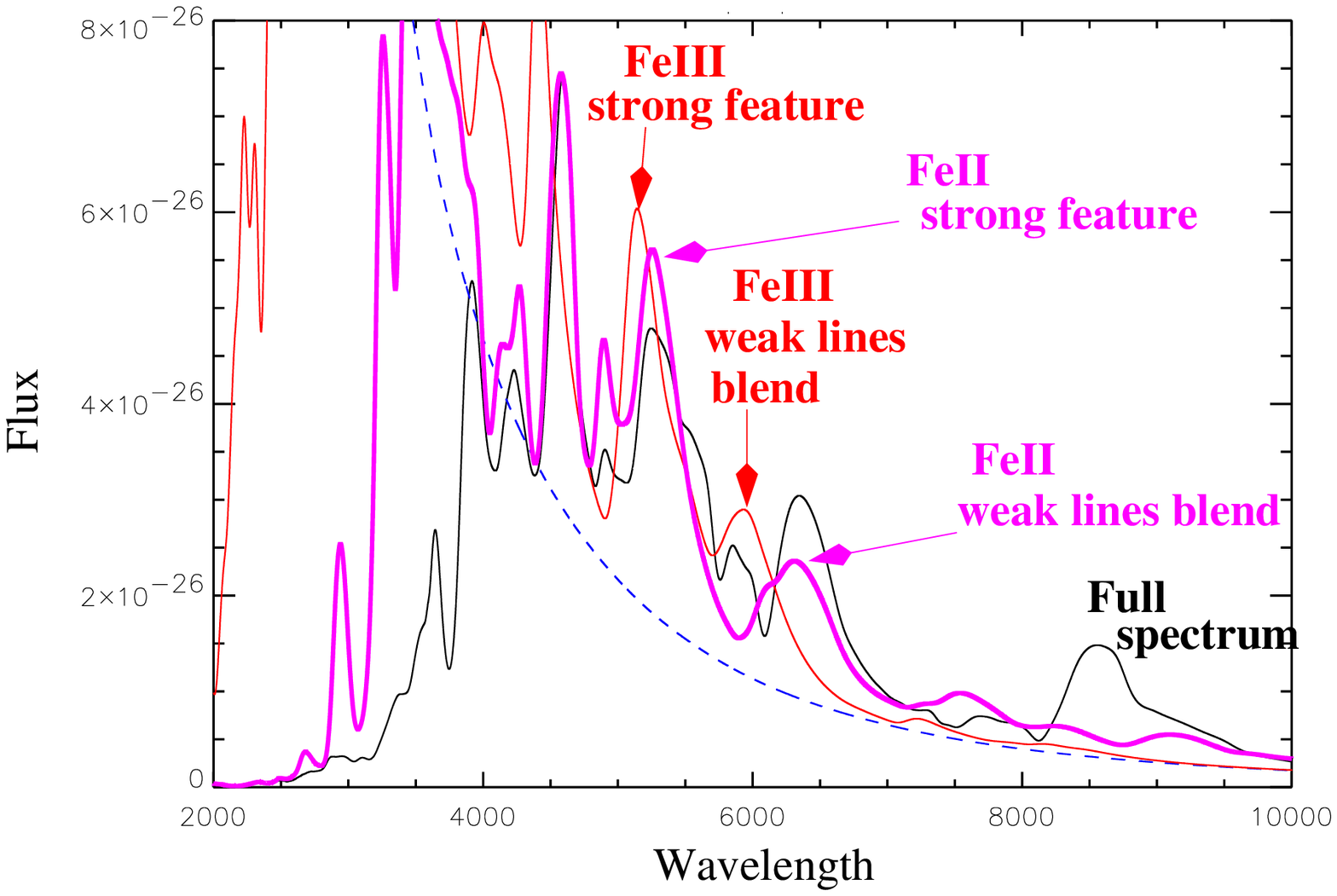}\\
  \includegraphics[width=0.48\textwidth]{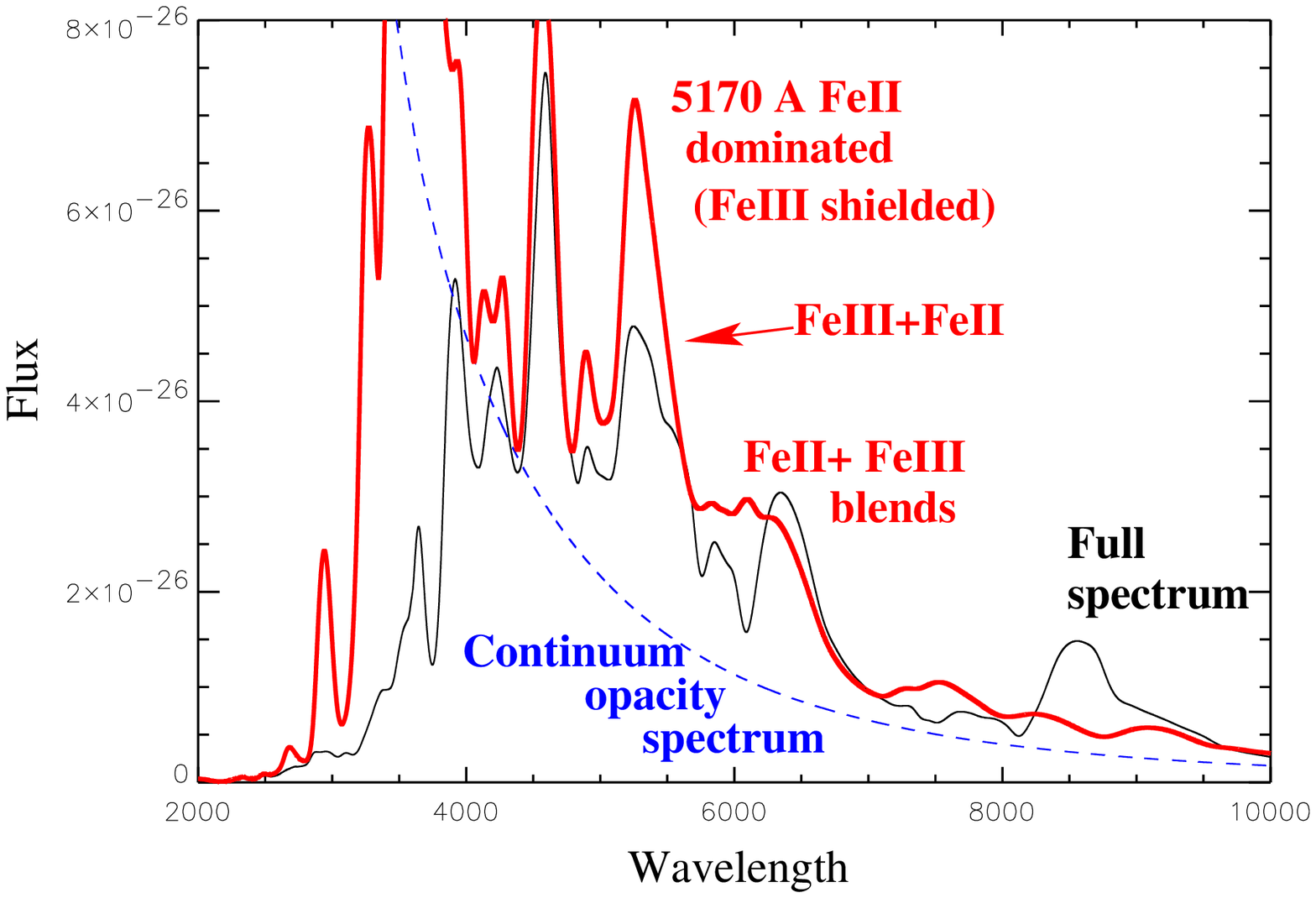}
  \caption{\FeII\ and \FeIII\ single ion spectra for our faintest
    simulated spectra, 20 days after explosion. The top panel is
    \FeIII\ in red and \FeII\ in pink. The bottom panel is \FeII\ +
    \FeIII\ two-ion spectra in red. In both cases the black
    spectrum is the full simulated spectrum and the blue spectrum is
    the continuum-only spectrum. }
  \label{fig:FeIIFeIIIonly}
\end{figure}

\begin{figure}[ht]
 \centering
 \includegraphics[width=0.48\textwidth]{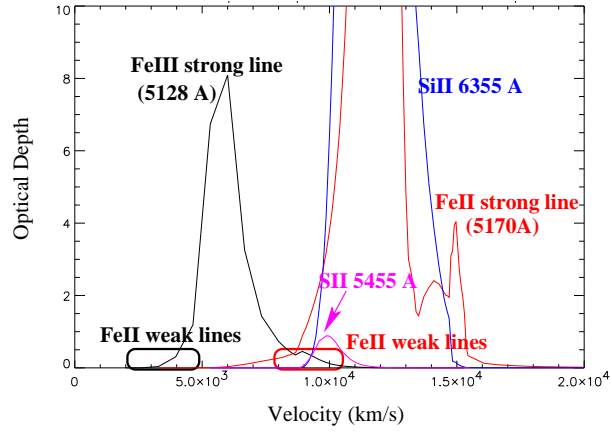}
 \caption{Optical depth of the most important lines of the \RSi\
   wavelength region at day 20. Weak lines always form deeper than the strong
   ones for \FeII\ and \FeIII. \FeIII\ $\lambda$5128 and \FeII\
   $\lambda$ 5170 are displayed as typical examples of strong \FeIII\
   and \FeII\ lines in the \RSi\ wavelength region.}
 \label{fig:tauFeII}
\end{figure}

In Fig.~\ref{fig:FeIIFeIIIonly} we display the \FeII, \FeIII\ and
\FeII +\FeIII\ two-ion spectra of our faintest
model at day 20. The \RSi\ wavelength region ($5000-7000$~\AA) gives us some
insight into the spectral formation process because of the smaller
number of strong lines than  in the bluer spectral
regions. The \FeIII\ and  
the \FeII\ peaks at 5900~\AA\ and 6100~\AA\ are blends of weak
lines formed at velocities of $\sim$ 5000 and
$\sim$ 9000~\kms, respectively. It follows from the optical depths
shown in 
Fig.~\ref{fig:tauFeII} that the 5128~\AA\ \FeIII\ peak and the
5170~\AA\ \FeII\ peak arise from strong lines formed at
velocities  of $\sim$ 7000 and $\sim$ 15000~\kms, respectively.   

Fig.~\ref{fig:FeIIFeIIIonly} shows how \FeIII\
and \FeII\ lines blend together. The strong \FeIII\ 5128~\AA\ feature is
shielded by the strong \FeII\ 5170~\AA\ one, and the two blends above
5900~\AA\ merge together. 

\clearpage

In the photospheric model the spectrum is considered to be a blackbody
continuum emitted by the photosphere on top of which strong, 
well-separated lines of ions present in the line
forming 
region add structure mainly by scattering light.
We define the ``pseudo-continuum'' as the spectrum formed by
weak line blends which produces more structure than a blackbody
spectrum. The photosphere is replaced in this picture by
a ``pseudo-photosphere'' forming at depth depending on
wavelength. This concept is widely used in the radiative transfer community.

This ``pseudo-photosphere'' is clearly seen in the results of
\citet{branchcomp105} who were able to model the spectrum of the
normal SN~Ia 1994D using \synow\ at 115 days past maximum
light, at least in the blue. Specifically, the results of
\citet{branchcomp105} show that even at late times (thus in the
deepest layers) the spectrum is dominated by permitted lines that can
be treated in resonant scattering. 

Similarly, the line forming region of the pseudo-photospheric model
does not have a fixed depth. We define it as the region where the
strong lines that create the final structure of the spectrum form.
In this picture, the strong  \FeIII\ 5128~\AA\ feature does not belong
to the line forming region because it forms in the same physical
region as the weak \FeII\ lines of the pseudo-photosphere with which
it blends, as can be seen in Fig. \ref{fig:tauFeII}. It adds
structure to the pseudo-continuum and is shielded by the strong
\FeII\ 5170~\AA\ feature forming further out. 
The existence of strong lines that form deep and of weaker lines that
imprint their features further out in the atmosphere leads to a
convergence between the pseudo-photospheric region and the line
forming region, making it almost impossible to separate them into two
clearly distinct physical regions.

The \FeII+\FeIII\ spectrum we display in Fig.~\ref{fig:FeIIFeIIIonly}
can be considered to be the pseudo-continuum spectrum in the
$5000-7000$~\AA\ region. It is formed by a large peak due to  
\FeIII\ $\lambda$5128 and neighboring strong \FeII\ lines,
followed by a flat region forming between 5000 
and 9000~\kms\ depending on wavelength. This  shows that
by construction \RSi\ couples spectral regions that form at very
different depths. 

In the ``pseudo-photospheric'' model, the flux transfer from the blue
toward the red is dominated by \FeIII\ and \FeII\ weak line
blends. Of course, weak cobalt and nickel lines also contribute,
but the study of their effects will be postponed to a later paper in
which the more complicated spectral formation in the $\lambda <
5000$~\AA\ region will be discussed. The colors of \SNeIa\ are therefore
dominated by weak lines of the heavy elements whose blending shapes
the pseudo-photosphere spectrum.

\section{Iron lines and colors}

\begin{figure}[ht]
  \centering
\includegraphics[width=0.45\textwidth]{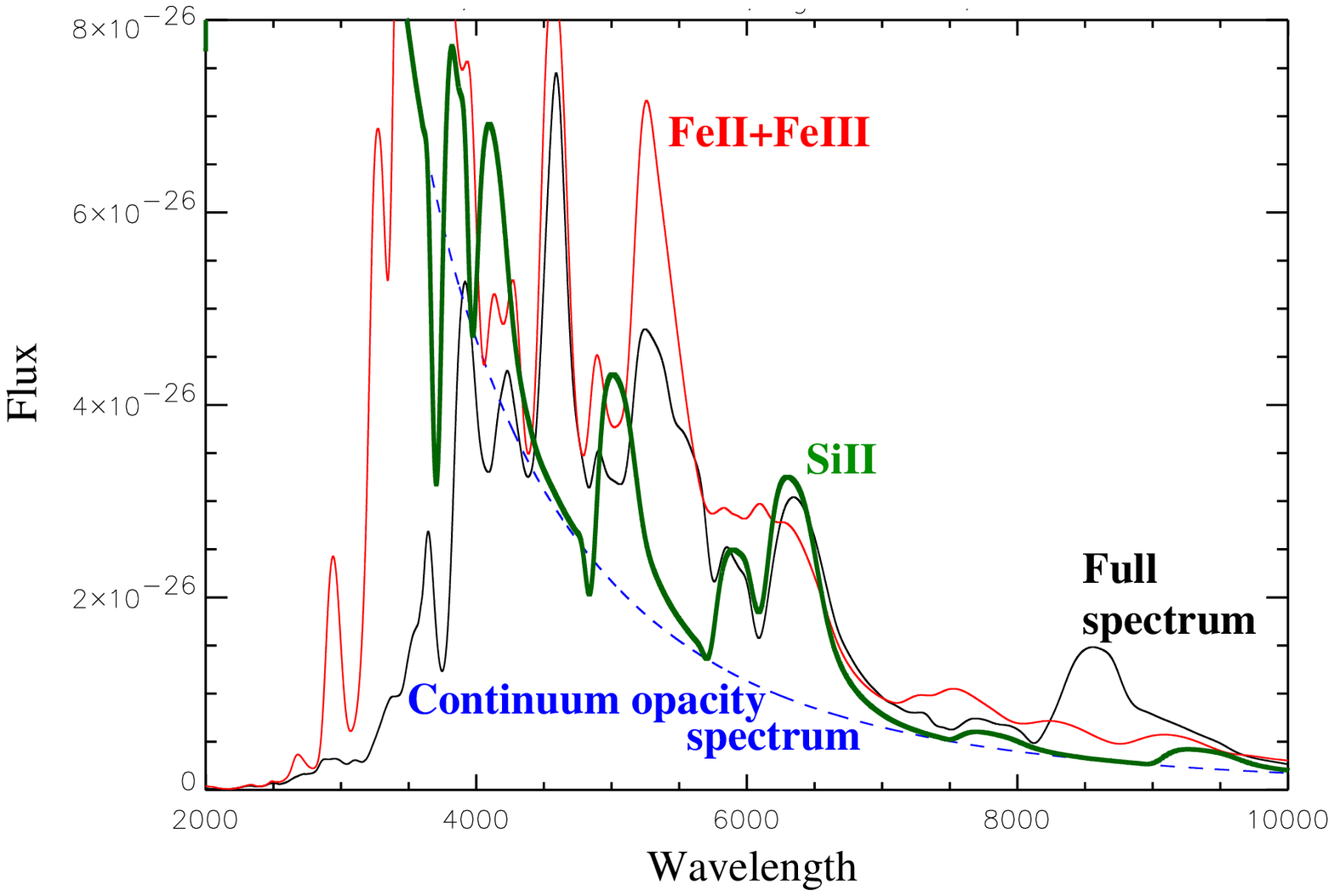}\\
  \includegraphics[width=0.45\textwidth]{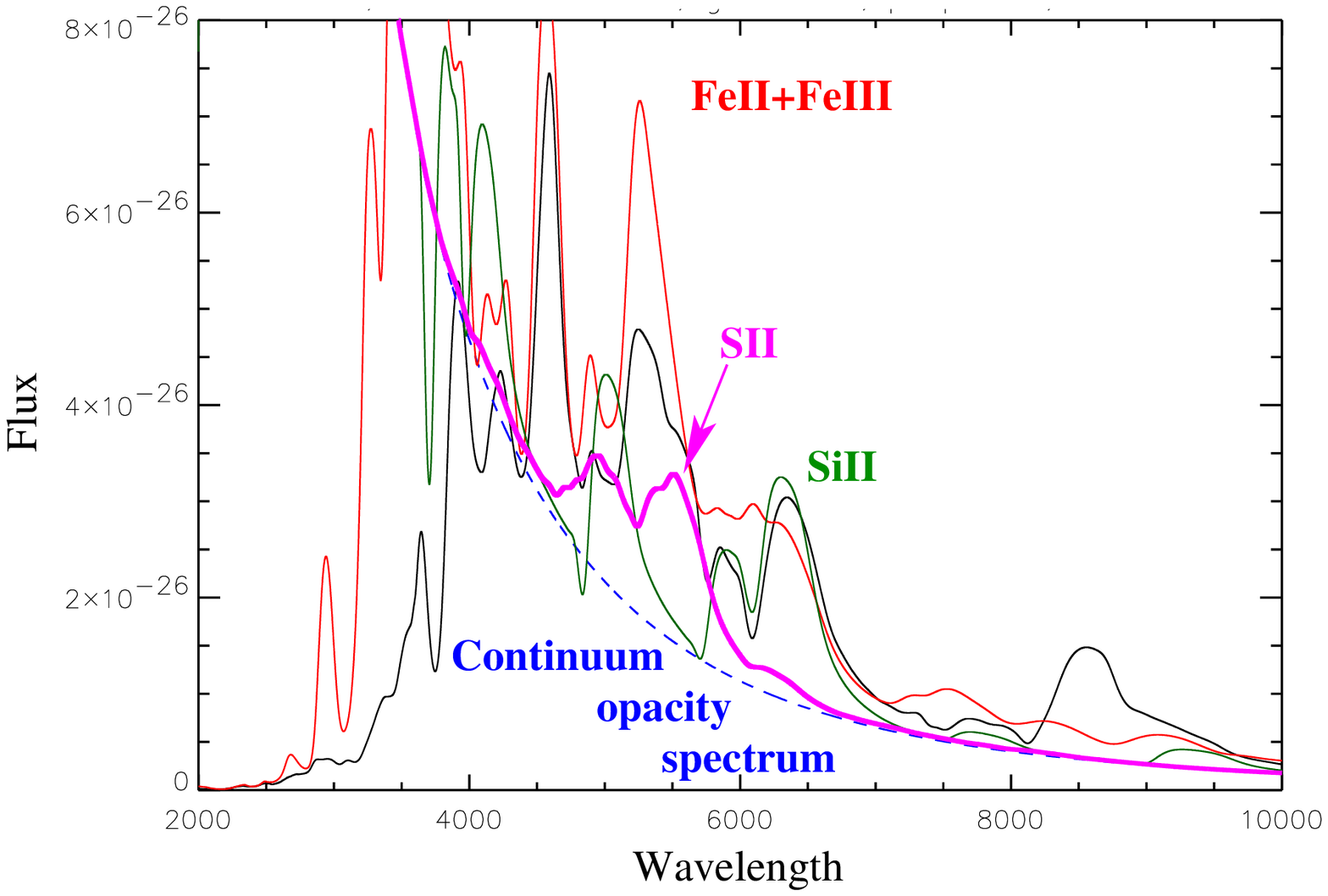}
\caption{\FeII\ (top panel) and  \FeIII\ (bottom panel) day 20 single ion spectra as a function of
  $M_{bol}$. Each spectrum is rescaled to the bolometric flux of the
  corresponding full synthetic spectrum. Since \FeIII\ forms deeper it is
  less sensitive to a change in the bolometric luminosity. \FeII\ lines are
  more efficient at transferring flux from blue to red. 
}
\label{fig:justfe_evol}
\end{figure}

\begin{figure}[ht]
  \centering
  \includegraphics[width=0.45\textwidth]{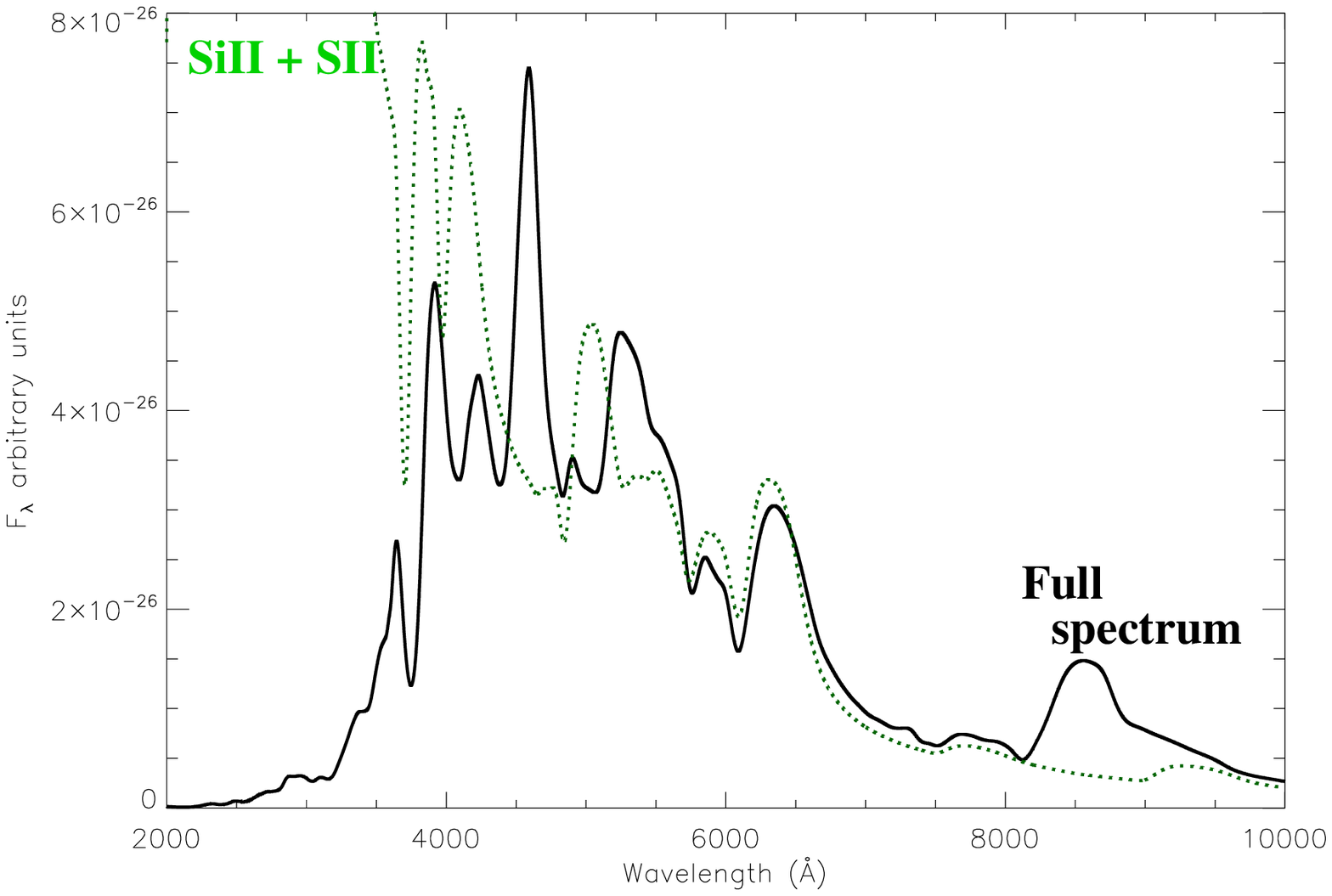}\\
  \includegraphics[width=0.45\textwidth]{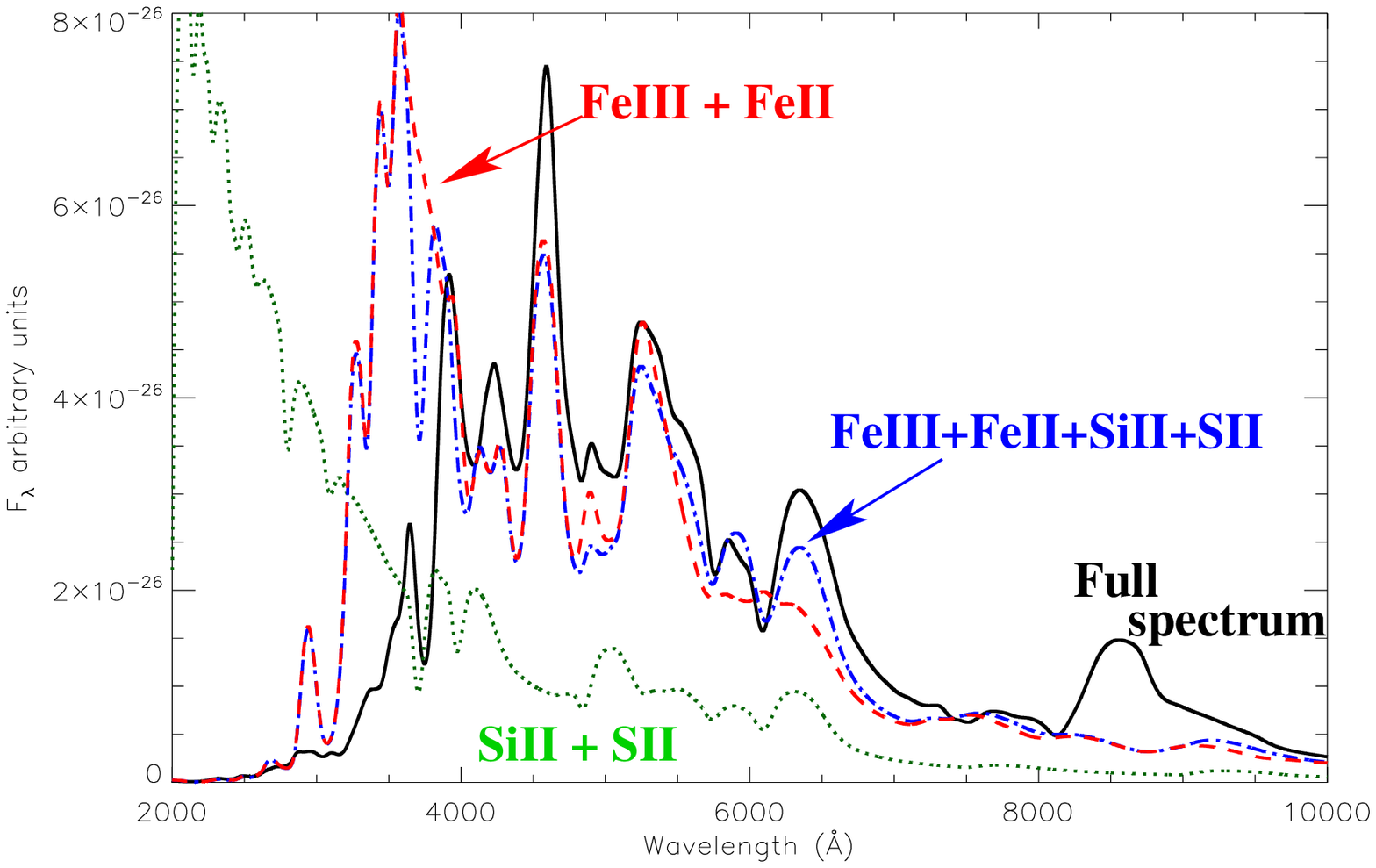}
  \caption{\FeII\ + \FeIII\  (top panel) and \SiII\ + \SII\
    (bottom panel) 
    two-ion day 20 spectra as a function of $M_{bol}$. The
    flux transfer from the blue to the red is dominated by the iron
    (and other iron peak elements in the full spectrum). All spectra
    are rescaled to have the same flux as the corresponding full
    synthetic spectra.} 
\label{fig:RSiSFeSi}
\end{figure}

\begin{figure}[ht]
  \centering
  \includegraphics[width=0.45\textwidth]{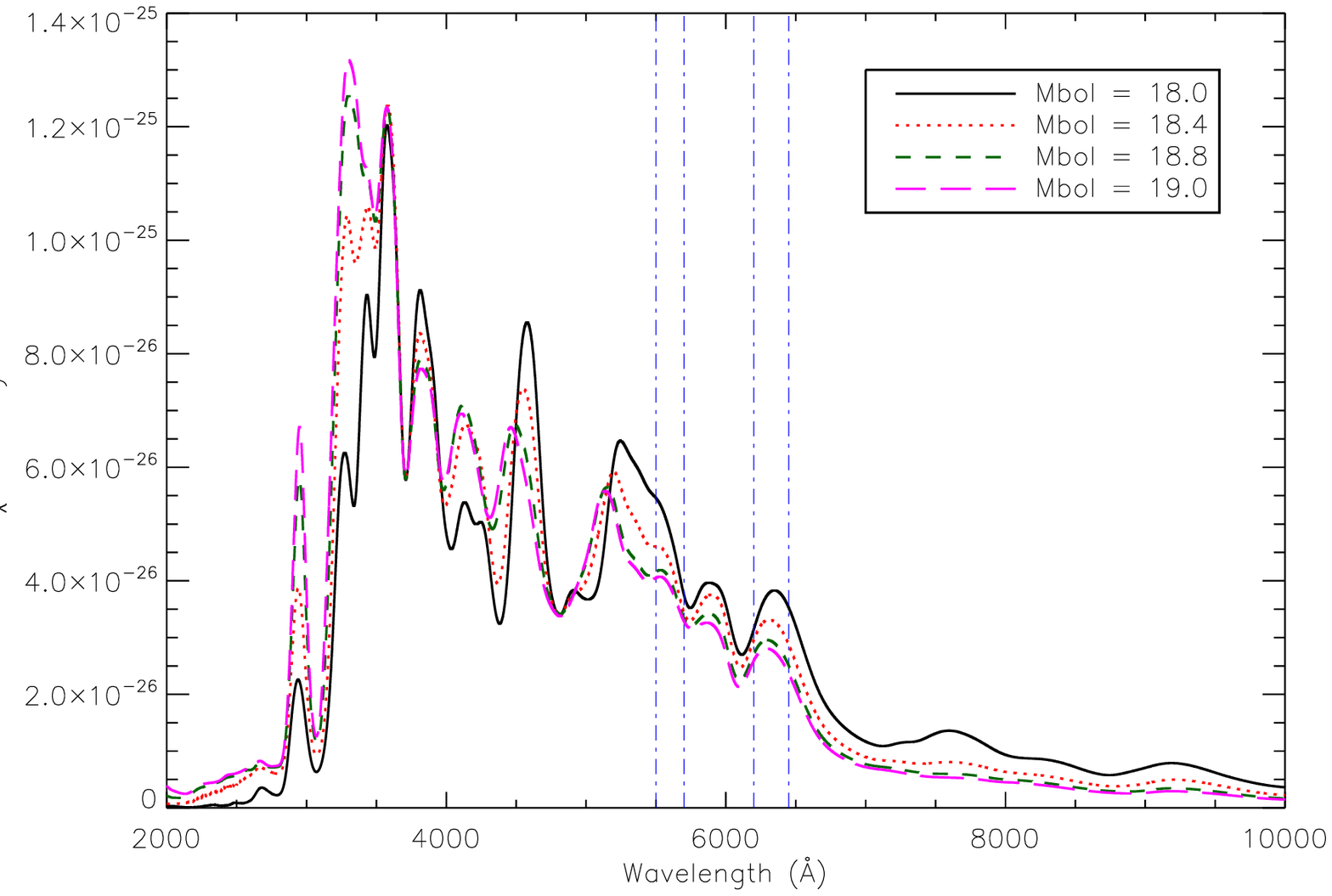}
\caption{\FeIII+\FeII+\SiII+\SII\ four-ion day 20 spectra as a
  function of $M_{bol}$. 
}
\label{fig:allfour}
\end{figure}

As the bolometric luminosity of our models is increased, the
temperature also rises, ionizing \FeII\ to \FeIII. The \FeII\
number abundance depletion decreases the 5170~\AA\ feature strength,
leaving \SII\ the dominant ion of the blue edge of the \RSi\
feature at $\sim$ 5455~\AA. 
This effect can be seen in
Figs.~\ref{fig:justfe_evol}--\ref{fig:allfour} where the \FeIII,
\FeII, \SiII, and \SII\ 
spectrum changes with bolometric luminosity are
displayed. The \FeII\ 5170~\AA\ feature strength decreases as \FeII\ is
ionized into \FeIII, unshielding the strong \FeIII\ 5128~\AA\ 
feature. Simultaneously, the flux transfer towards the red 
decreases because \FeII\ is much more efficient at transferring flux
toward the red than \FeIII\ as can be seen in
Fig.~\ref{fig:FeIIFeIIIonly}. This efficient flux transfer from
the blue to the red is due to numerous blue and UV \FeII\ lines that
amount to a large opacity because of line blanketing.

\clearpage
\begin{figure}[ht]
  \centering
\includegraphics[width=0.8\textwidth]{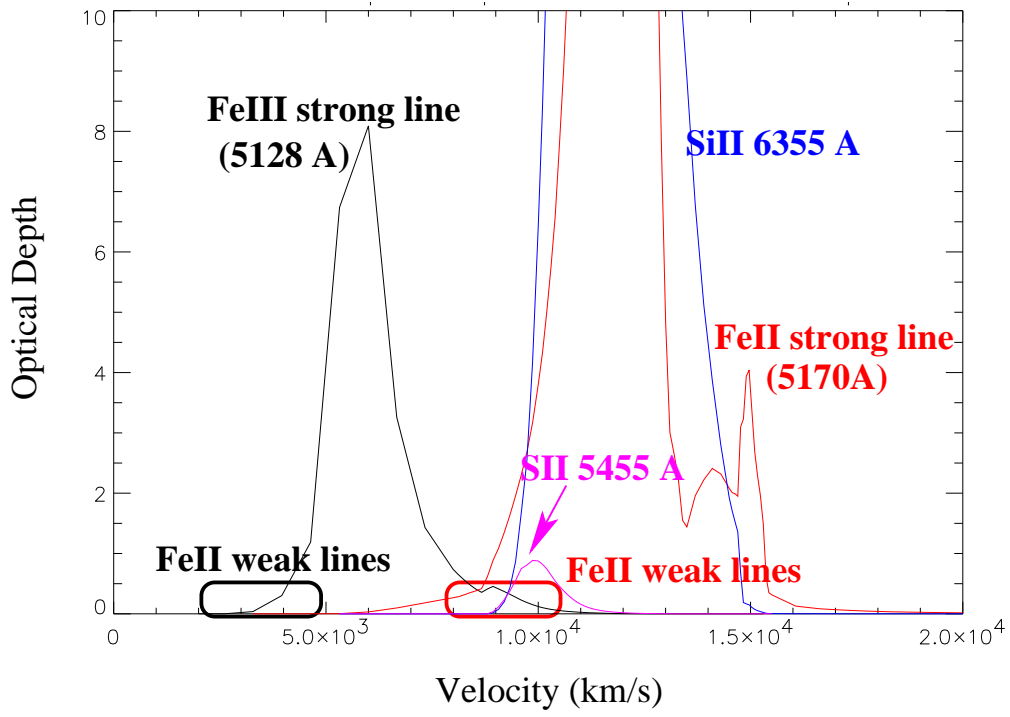}\\
\caption{$B-V$ color as a function of 
  $M_{bol}$ for the different epochs in our grid. 
}
\label{fig:bmvevolution}
\end{figure}

\clearpage

\clearpage

Figure \ref{fig:FeIIFeIIIonly} also shows that in the 5000-8000~\AA\
wavelength region, \FeIII\ features remain very stable over the range
of bolometric luminosities explored, while \FeII\ features vary much
more. This is not surprising, as \FeIII\ lines form much deeper, in a
region that is hotter and denser, where physical conditions vary less than in
the outer regions.

We see in Figure \ref{fig:RSiSFeSi} (top panel) that the \FeIII+\FeII\
two-ion 
spectrum undergoes a strong  increase of flux below 5500~\AA,
while it remains very stable above, when the bolometric luminosity
increases. This translates into a strong variation in color of the
full synthetic spectrum. 
On the other hand, the \SiII+\SII\ two-ion spectra
displayed in Fig.~\ref{fig:RSiSFeSi} (bottom panel) remain very consistent in 
shape, while their flux follows the continuum-only spectrum flux increase.  

The flux variation displayed in Fig.~\ref{fig:allfour} is thus
dominated by the ``pseudo-continuum'' of \FeIII\ and \FeII\ lines,
especially the weak line blends, on top of which the strong \SiII\
and \SII\ lines add structure. The flux transferred by the well
separated lines formed in the line forming region is small compared to
the flux transferred at depth by iron lines, but is by  no means negligible
when using SNe~Ia for precision cosmology. It is therefore crucial to
have a good description of the full spectrum in order to perform
K-corrections.

Figure \ref{fig:full_spect_tempevolution} shows very close
agreement between the full synthetic spectrum and the
\FeIII+\FeII+\SiII+\SII\ four-ion spectrum above 5000~\AA,
proving that in the W7 model, the other metallic lines have a secondary
importance in transferring flux compared to \FeIII\ and \FeII\ lines.

Faint \SNeIa\ are expected to be redder because of a larger
proportion of \FeII\ (as compared to \FeIII) and not because the
underlying na\"ive blackbody is colder. That is, accepting that the
primary variation for the peak luminosity is due to nickel mass
\citep{nugseq95}, dimmer supernovae are redder because the underlying
ionization stage is primarily \FeII. Figure~\ref{fig:bmvevolution}
clearly displays this brighter-bluer evolution for each epoch we
simulated. 

While the variation with epoch is beyond the scope of this
work, the variation with luminosity in part mimics the variation with
epoch to some extent. Moreover, the spectra at 10 days after explosion
are much bluer than the later ones, because the supernova is much
denser and hotter at this epoch, leading to a higher degree of
ionisation of the iron core.

This \FeIII-\FeII\ effect on flux transfer agrees with the 
analysis of \citet{kasen06b}, which explains the  second red bump in
\SNeIa\ light curves by the change in opacity due to the recombination
of \FeIII\ into \FeII: when the temperature decrease in the \SNeIa\
envelope is enough for \FeIII\ to recombine into \FeII\ more flux is
transferred from the blue into the red, causing the second red bump.   

\subsection{Silicon and Sulfur lines}

\begin{figure}[ht]
  \centering
  \includegraphics[width=0.48\textwidth]{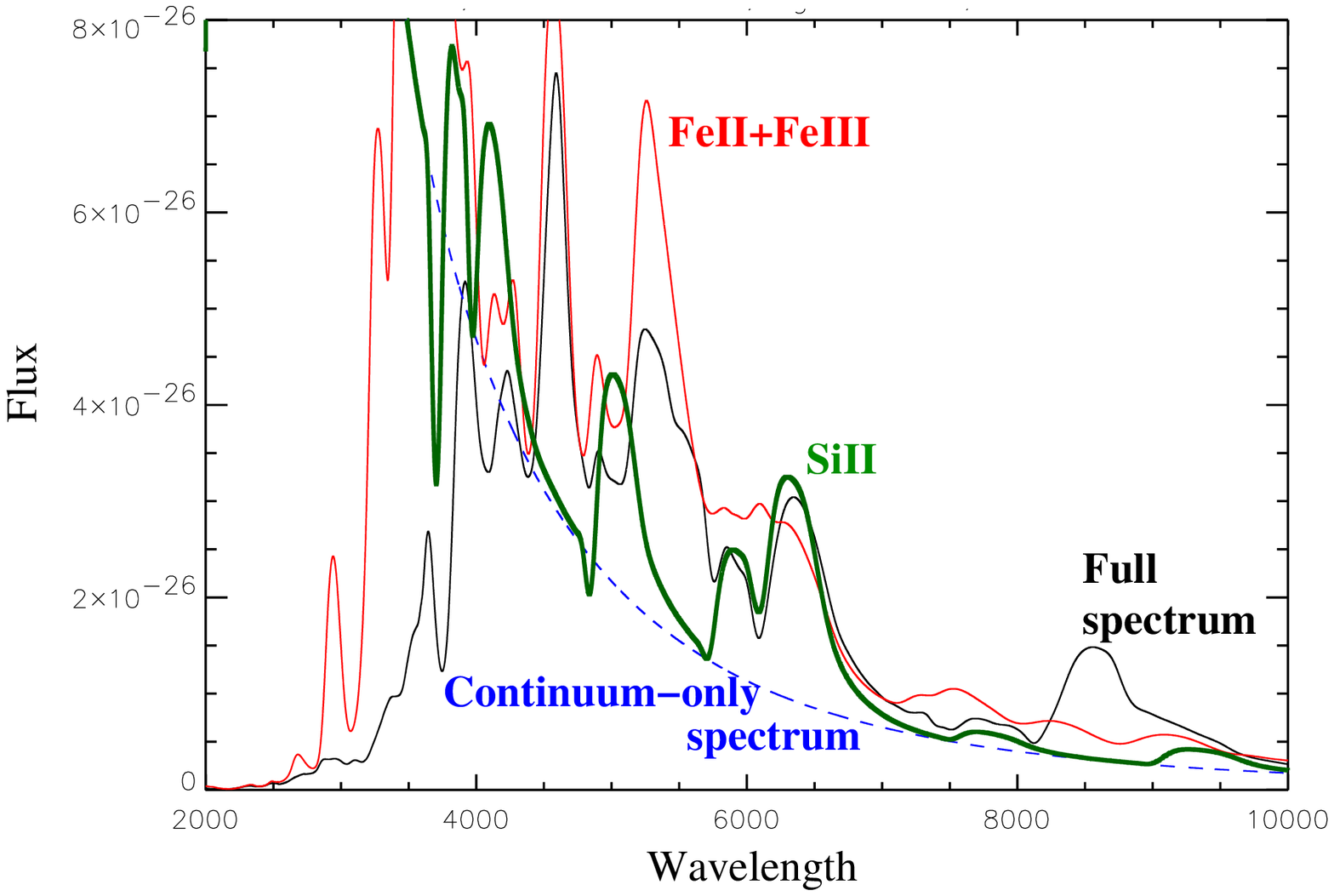}\\
  \includegraphics[width=0.48\textwidth]{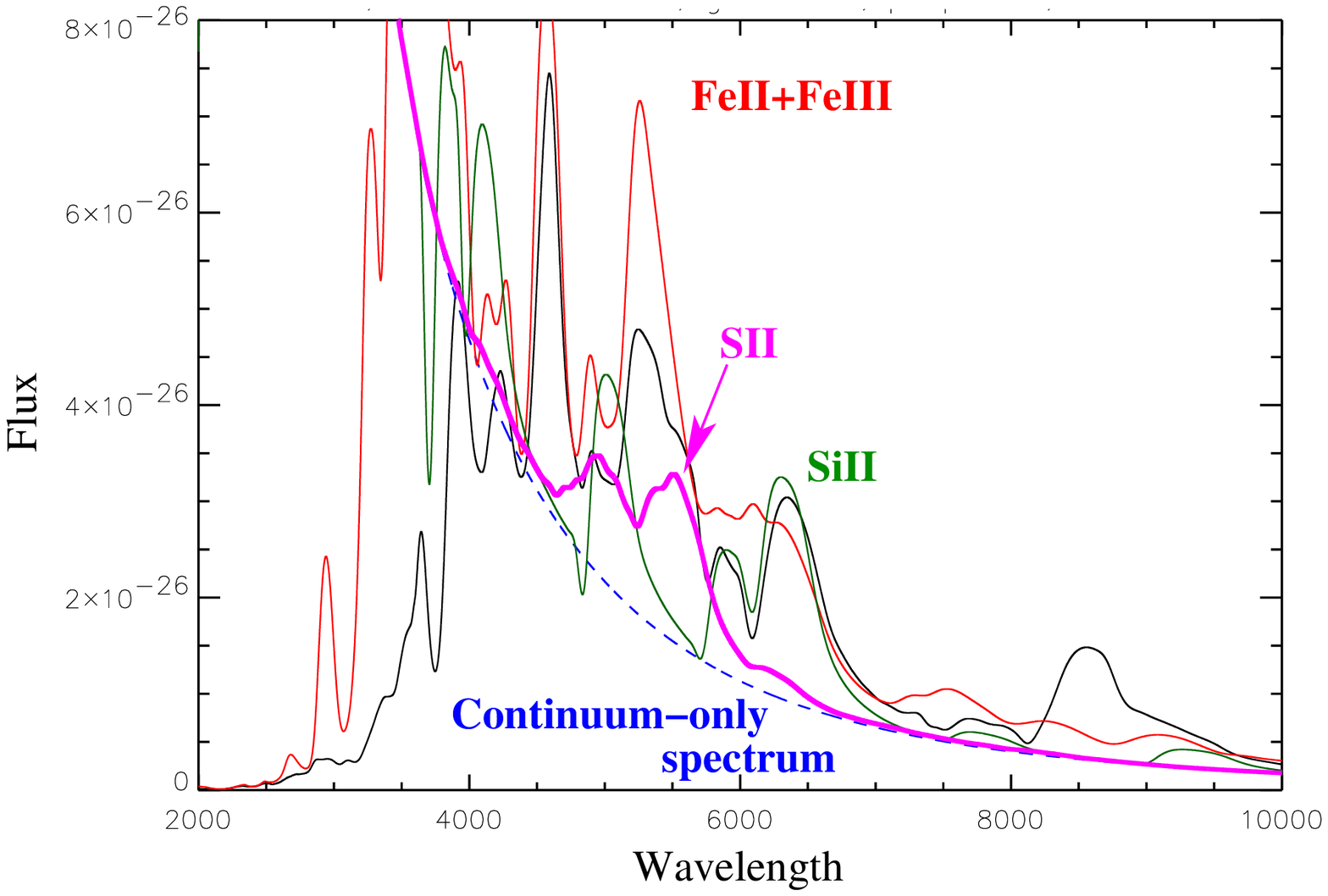}
  \caption{Single ion spectra for our faintest luminosity 20 days
    after explosion. The top panel shows spectra without \SII, the
    bottom panel includes \SII. They show that \SiII\ lines alone 
    can not account for the \RSi\ wavelength 
  region shape. \FeII\ and \FeIII\ contribute with their weak lines to
the ``pseudo-continuum''. The blue edge ($<$5600~\AA) of the \RSi\
region is dominated by \SII\ lines. }
  \label{fig:SiIIonly}
\end{figure}

Fig.~\ref{fig:SiIIonly} (top panel) displays \FeII +\FeIII\ two-ion spectrum
and \SiII\ single ion spectrum. \SiII\ lines form in the line
forming region, contributing little to the global flux transfer but 
adding lots of structure to the spectrum. 
As previously mentioned, the $\sim 6100$~\AA\ trough as well as the
\SiII\ \SiIIred\ emission peak are dominated by a strong \SiII\ line. The
\SiII\ \SiIIblue\ line is also seen to dominate the second ``P-Cygni''
trough, but \SiII\ alone fails to account for the blue edge at
$\sim 5600$~\AA. We already ruled out \TiII, and the \FeII\ 5170~\AA\
peak is too blue to account for this edge.  
Fig.~\ref{fig:SiIIonly} (bottom panel) displays the
\SII\ single ion spectrum in addition to the previous ones,
showing that \SII\ lines can account for the bluer edge of the \RSi\
wavelength region. 
\clearpage
\clearpage

\begin{figure}[ht]
  \centering
  \includegraphics[width=0.48\textwidth]{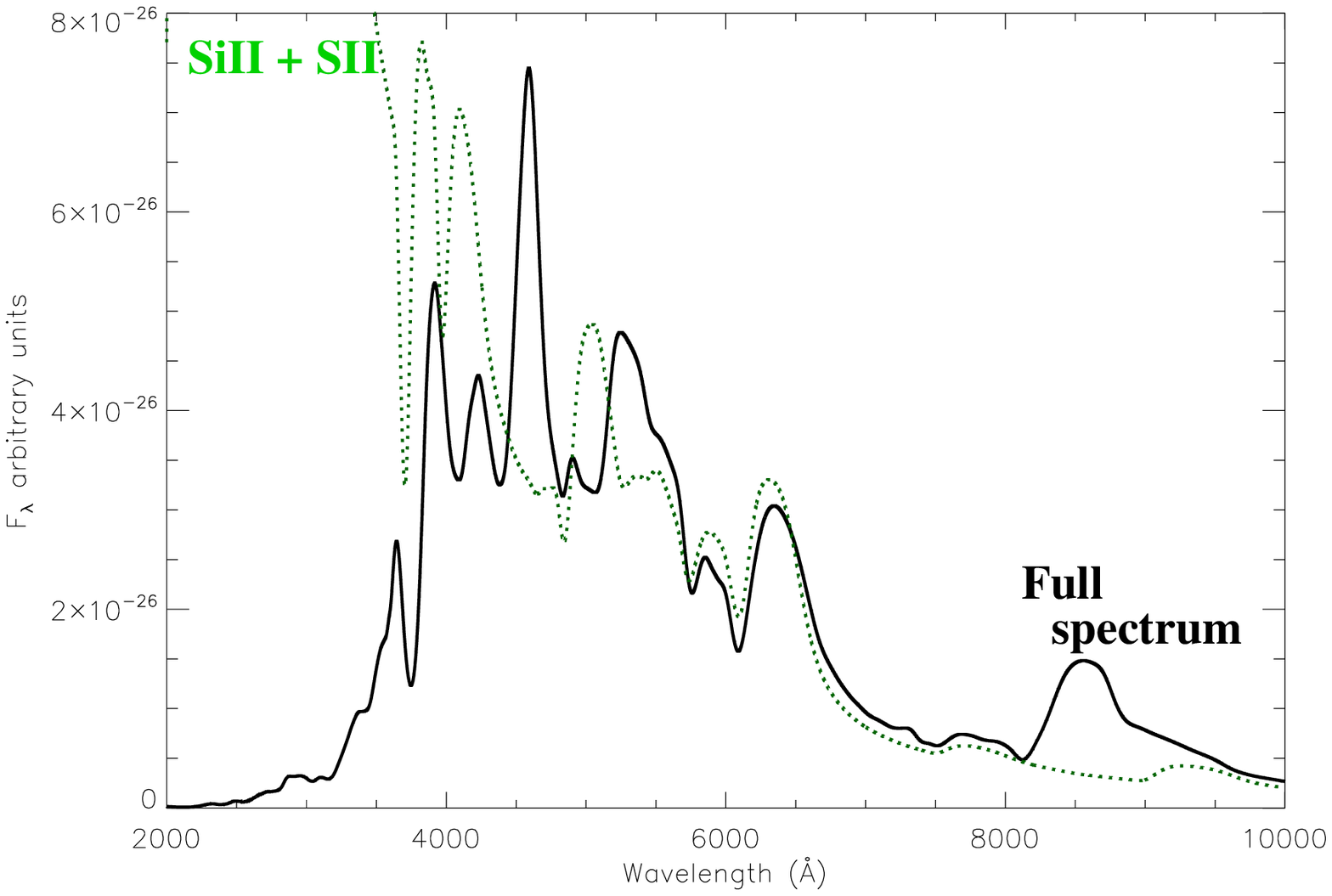}\\
  \includegraphics[width=0.48\textwidth]{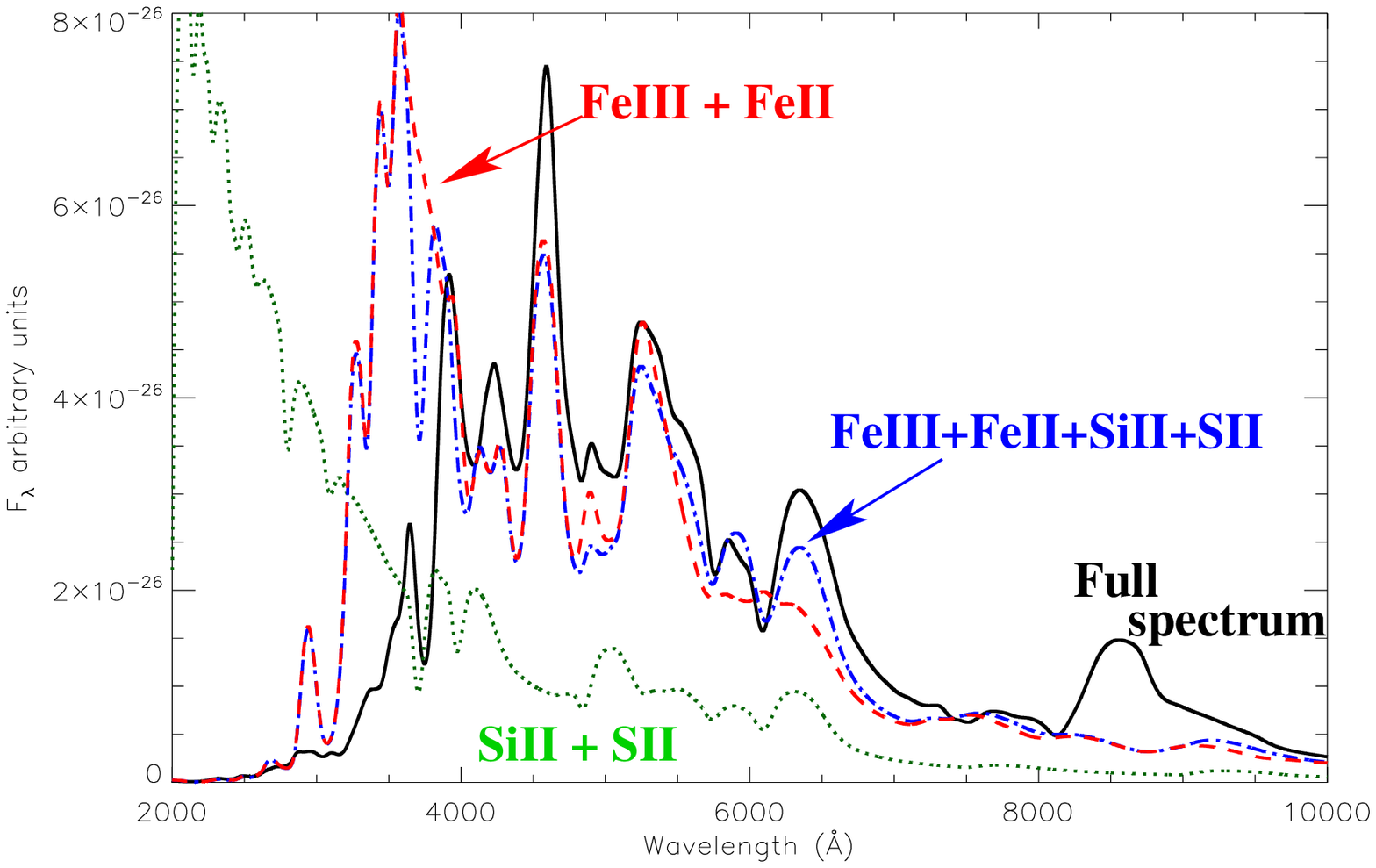}
  \caption{\SiII+\SII\ two-ion day 20 spectra (top panel, not rescaled,
    bottom panel, rescaled)
    } 
  \label{fig:SIIFeIIFeIII}
\end{figure}

The \SiII+\SII\ spectrum of Fig.~\ref{fig:SIIFeIIFeIII} (top panel) 
shows that the relative strengths of the peaks and depths due to
\SiII\ and \SII\ in the \RSi\ wavelength region are consistent with 
the spectrum of the full \nomw\ model. In Fig.~\ref{fig:SIIFeIIFeIII}
(bottom panel), 
where all the spectra are rescaled to have the same total
flux, if the \RSi\ line features are dominated by
\SiII\ and \SII, it is the \FeIII\ and \FeII\ lines that provide the
flux background.  
The \FeIII+\FeII+\SiII+\SII\ spectrum shows that between 5000~\AA\ 
and 8000~\AA\ the spectrum of the full \nomw\ model is well reproduced by 
these 4 ions alone.  This is an important result for the
understanding of the theoretical explanation of the correlation
between the line ratios and the absolute $B$ magnitude. Since we have
shown that the \RSi\ ratio, that is the flux in this wavelength range,
depends only on the iron,  silicon, and sulfur; to correctly model the
line ratio, one must only have a model that correctly accounts for
these ions. Clearly we are not there yet, but this result
significantly reduces the parameter space that must be studied to
obtain physical understanding. 

However, we must be cautious due to the fact that our models do not
reproduce the observed spectra. For example, W7 does not reproduce the
shape of the Si~II 6100\ang\ feature \citep{bbbh06} and thus we could
be missing at least one important species in our calculations. Turning
this around shows that obtaining the correct variation of \RSiS\ with
luminosity could be an important filter for obtaining physically
correct models. 

Similarly, the strong iron features around 4500~\AA\ and 5170~\AA\ are
unrealistically large compared to observed spectra. These strong
features are a finer probe of a model quality than the overall
shape of the spectrum: the thick blend of lines of the iron core is
much less sensitive to an abundance variation than the strong lines,
forming where the iron blanket has become optically thin. 

\clearpage
\clearpage

\section{\RSiS: observational support of the ``multi-layered spectral
  formation''} 

\subsection{\RSiS\ predicted trend: a hotter \phx\ spectrum\label{rsisdef}}

We defined a line ratio \RSiS\ \citep{bongard06a} as 
\begin{equation}
  \mbox{\RSiSS}=\frac{\int_{5500}^{5700}F_{\lambda}d\lambda}{\int_{6450}^{6200}F_{\lambda}d\lambda}.    
  \label{RSiSdef}
\end{equation}
We showed that it was better correlated to luminosity than \RSi, and
deferred the motivation to the present paper. Here is
where we address this question.

\begin{figure}[ht]
\centering
  \includegraphics[width = 0.8\textwidth]{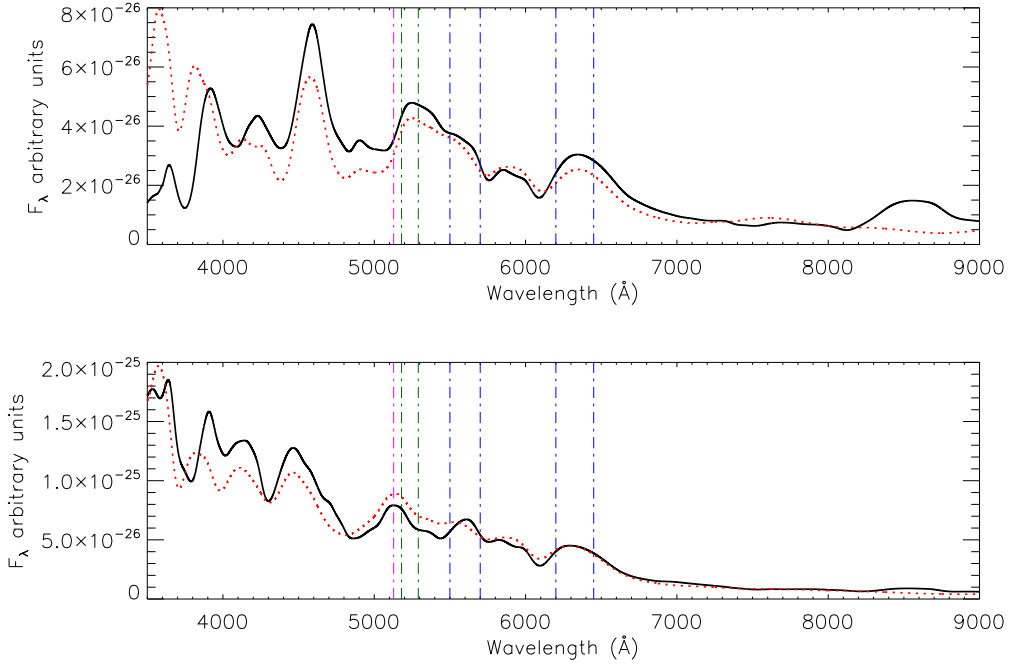}
  \caption{The full synthetic day 20 spectrum (solid line) is compared to the
    day 20 four-ion spectra including \FeIII, \FeII, \SiII,
    and \SII\ (dotted line).  The top panel shows $M_{bol}= -18.0$, the
    bottom panel 
    $M_{bol}= -19.0$. The pink vertical line indicates  \FeIII\ 
    $5128$~\AA\ feature, while the green vertical region shows \FeII\
    strong feature $5170-5291$~\AA\ that dominates at lower
    luminosities. The blue vertical zones are zones where \RSiS\  is
    defined.  
}
  \label{fig:full_spect_tempevolution}
\end{figure}

As shown in the previous sections, the \RSi\ wavelength region is
dominated by \FeIII, \FeII, \SiII, and \SII\ lines. Moreover, for
luminosities corresponding to ``normal'' \SNeIa, the bluer edge
of this wavelength region is dominated by \SII, since the envelope is
still too hot to allow the strong \FeII\ $5170$~\AA\ feature to dominate. As
the luminosity decreases, the strength of the \FeII\ $5170$~\AA\ 
feature increases, this behavior is illustrated in
Fig.~\ref{fig:full_spect_tempevolution}. This effect will have 
differing importance for the two different definitions of the line
ratio, \RSi\ and \RSiS. \RSi\ will hardly be affected since the depth of
the blue \SiII\ feature does not vary significantly; however, \RSiS\
uses just this wavelength region and will thus be strongly impacted by
the growth of the \FeII\ $\lambda5170$ line.

The  ``multi-layered line formation''  picture shows that the depth of
the  troughs are not the pertinent quantities to probe the temperature
sequence, since the ``absorption'' features are formed by a complex
blend of absorptions and emissions due to ions spanning a wide
range of depth. The intermediate $\sim 5900$~\AA\ \RSi\  peak has
the same problem, as it is a blend of  \FeIII, \FeII, and \SiII\ lines.   
A direct corollary is that \RSi\ does not measure line strength
evolution as is commonly believed. 

\clearpage
\clearpage

The motivation for the definition of \RSiS\ was to isolate contributions from
\SiII\ and \SII\ lines forming in the same physical region between
10000~\kms\ and 15000~\kms. 

There are at least two options for the correlation of \RSiS\ with
absolute blue magnitude:  the silicon and sulfur lines form in the
same physical region and their strength is simply determined by the
local conditions, or more likely \RSiS\ does not measure the region where
silicon and sulfur lines form, but rather the conditions where the
iron ``pseudo-continuum'' forms.

Keeping in mind that one should be careful when interpreting single
ion spectra, Fig.~\ref{fig:RSiSFeSi} supports this latter explanation.
Fig.~\ref{fig:RSiSFeSi} (top panel) shows that the \FeII\ and \FeIII\
blend varies little in the 6100~\AA\ wavelength zone used to
compute \RSiS. The consistency in the \SiII+\SII\ two-ion
spectra shape displayed in Fig.~\ref{fig:RSiSFeSi} (bottom panel) suggests a
small variation of the relative strength of the \SiII\ and \SII\
features. We therefore favor the hypothesis that \RSiS\ traces the
evolution in shape of the ``pseudo-continuum'' dominated by iron
lines. As such, it should be better correlated with luminosity, since
it probes the deeper layers of the supernova. On the other hand, it
will be more sensitive to pollution by lines forming further out,
especially strong \FeII\ lines.

\subsection{Synthetic vs observed spectra}

Up to now we have used the \nomw\ model as a reasonable physical explosion
model of \SNeIa, but we did not try to
match with observed spectra. We 
used the knowledge of the abundance structures and the physical
structure obtained by \phoenix\ to probe the line formation process,
considering the \nomw\ model as a theoretical supernova of which we
had complete knowledge. 

\begin{figure}[ht]
\centering
 \includegraphics[width=0.7\textwidth]{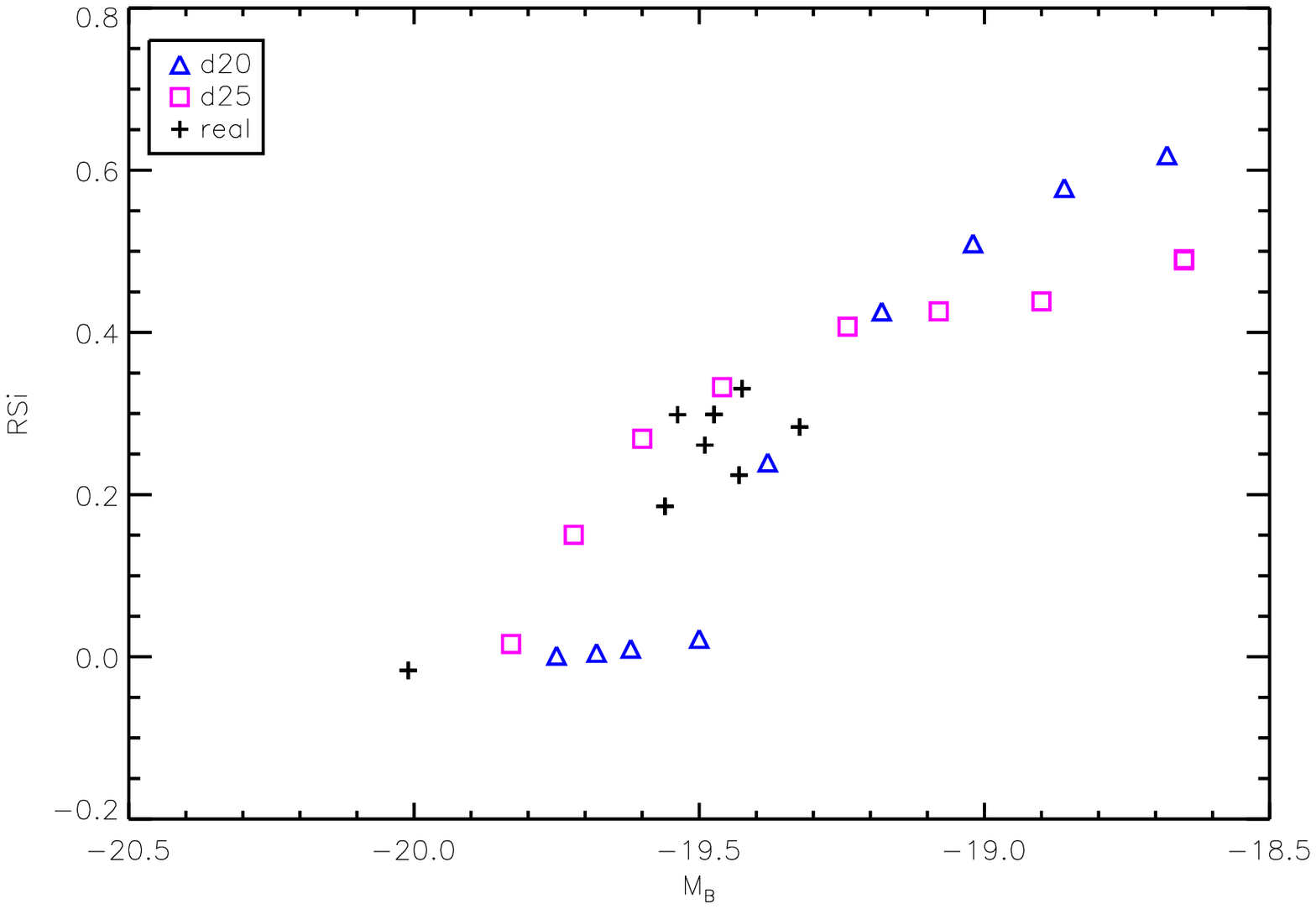}\label{fig:RSiSphx_a}\\
\includegraphics[width =0.7\textwidth]{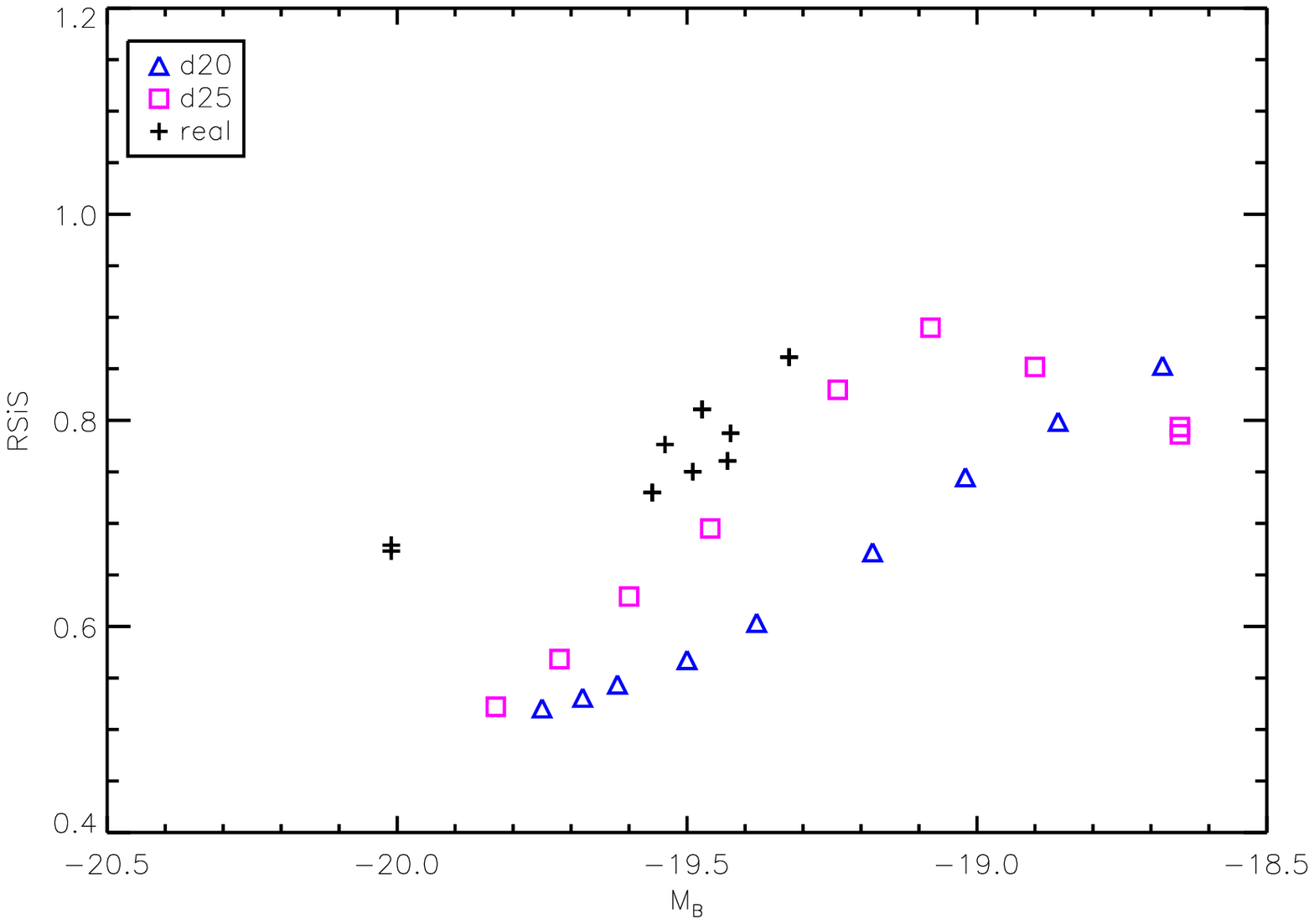}\label{fig:RSiSphx_b}
  \caption{\RSi\ vs $M_{B}$ (top panel) and \RSiS\ vs $M_{B}$ (bottom
    panel) The real
    data are displayed for reference, and are discussed thoroughly in
    \citet{bongard06a}. The trends of real and synthetic spectra for 20 and
    25 days after maximum are similar. The change in slope between day 20
    and day 25 days after explosion is due to \FeII\ lines starting to
    dominate the \RSi\ wavelength region.}
  \label{fig:RSiSphx}
\end{figure}

The \RSiS\ ratio correlates well with
absolute blue magnitude \citep{bongard06a}. In Fig.~\ref{fig:RSiSphx}
we display \RSi\ 
and \RSiS\ for our day 20 and day 25  models as well as
real \SNeIa\ data. The calculated \RSi\ appears to reproduce the
observed data quite well for both days 20 and 25. It follows the
``normals'' quite well and may even extend to the regime of very
low luminosity SNe~Ia such as SN 1991bg. At first glance, it would
appear that the calculated \RSiS\ fails to match the observations;
however, if one restricts one's attention to the ``normals'', 
[Fig.~\ref{fig:RSiSphx} (bottom panel)], in fact the agreement between
\RSiS\ (for day 25) and the data is better than that for \RSi. The
change in slope we observe around $M_{B} = -19.1$ for day 25 comes
from the strong \FeII\ 5170~\AA\ feature that dominates over \SII\ at
lower luminosities. 

It should be noted that the W7 model is not expected to reproduce SNe~Ia
light curves, and certainly cannot be expected to match the observed
diversity from fast to slow decliners.
Therefore, the time since explosion should be considered
only as indicative, and related only with great care to real
supernov\ae\ epochs.  

\clearpage
\clearpage

\section{Conclusion}

We used a grid of LTE \phoenix\ synthetic spectra using the \nomw\
model, to study the spectral formation in \SNeIa\ around the time of maximum
light. Instead of focusing on fitting synthetic to observed spectra,
we probed the detailed line formation in the synthetic spectrum to try
to understand which ion was responsible for which feature, while
taking detailed line blending into account.

We analyzed the synthetic spectra in detail to study the
line formation in the \RSi\ wavelength region, showing that the
\SiIIred\ peak and the 6100~\AA\ trough were dominated by \SiII\ lines
forming over a blend of  \FeII/\FeIII\ lines. The intermediate
5800~\AA\ peak has been shown to be a complex blend of  \SiII\
\SiIIred, \SiII\ \SiIIblue, \FeIII, and \FeII\ weak lines. The \TiII\
contribution to the 5500~\AA\ trough has been ruled out. We 
showed that the redder edge was dominated by the \SiIIblue\  \SiII\
line whereas the bluer one was dominated by \SiII\ 5455~\AA\ line or
\FeII\ 5170~\AA\ depending on luminosity. Based on our unraveling of
the line formation in this wavelength region, we are able to
illustrate our motivation for the definition of \RSiS,  
isolating the \SiII\ and \SII\ contributions of the \RSi\ region.   

We described a multi-layer model where the 
observed pseudo-continuum is formed throughout the entire supernova and
thus strong features from multiple ionization stages can occur in the
observed spectrum.
We stress the importance of the numerous weak iron
lines and show that their blends dominate the flux transfer. As a
corollary, in the multi-layer model, the inner spectrum is not close to
a Planck function but contains much more structure background. 
We showed that the ionization stage of the iron core
dominates the \SNIa\ colors, explaining the brighter-bluer relation at
maximum light. 
Even though a detailed study of time evolution is beyond
the scope of this paper, we showed that our results where also
qualitatively consistent in this regard, our earlier spectra being
systematically much bluer than our later ones.

\acknowledgements
  We thank the anonymous referee for comments which significantly
  improved the organization of this paper.
  This work was supported in part by by NASA grants NAG5-3505 and
  NAG5-12127,  NSF grants AST-0307323, AST-0506028, and
  AST-0707704, and US DOE Grant DE-FG02-07ER41517.  S.~Bongard and
  E.~Baron acknowledge support from the 
  US Department of Energy Scientific Discovery through Advanced
  Computing program under contract DE-FG02-06ER06-04.  
  This research used resources of the National Energy Research
  Scientific Computing Center (NERSC), which is supported by the
  Office of Science of the U.S.  Department of Energy under Contract
  No. DE-AC03-76SF00098; and the H\"ochstleistungs Rechenzentrum Nord
  (HLRN).  We thank all these institutions for a generous allocation
  of computer time.


\end{document}